\documentclass[12pt,preprint]{aastex}

\usepackage{graphicx}

\usepackage{psfig}

\newcommand{\Msunyr}{\mbox{$M_{\odot}~{\rm yr^{-1}}$}}

\newcommand{\my}{$\mu$m}

\def\bgam{\hbox{\rm B${\gamma}$}}
\def\hap{\hbox{\rm H${\alpha}$}}

\def\bap{\hbox{\rm B${\alpha}$}}

\def\hedubt{\hbox{\rm \HeI\ ${2.112~\mu}$m}}
\def\hedubi{\hbox{\rm \HeI\ ${2.112/3~\mu}$m}}
\def\hetrit{\hbox{\rm \HeI\ ${1.700~\mu}$m}}

\def\tc504{\hbox{$\tau_{\rm 504}$}}

\def\tauro{\hbox{$\tau_{\rm R}$}}

\def\taur23{ \hbox{$\tau$=2/3}}

\def\Vesc{\hbox{V$_{\rm es}$}}

\def\Mdot{\hbox{$\dot {\rm M}$}}
\def\Rsun{\hbox{R$_\odot$}}

\def\Rstar{\hbox{R$_*$}}

\def\Rro23{\hbox{R$_{\tauro=2/3}$}}
\def\R23{\hbox{R$_{2/3}$}}

\def\Lsun{\hbox{L$_\odot$}}
\def\Lstar{\hbox{L$_*$}}
\def\Msun{\hbox{M$_\odot$}}
\def\Msunyr{\hbox{M$_\odot\,$yr$^{-1}$}}

\def\Teff{\hbox{T$_{\rm eff}$}}

\def\T23{\hbox{T$_{2/3}$}}

\def\Vesc{\hbox{V$_{\rm esc}$}}
\def\Vinf{\hbox{$v_\infty$}}
\def\kms{\hbox{km$\,$s$^{-1}$}}
\def\kpc{\hbox{kpc}}

\def\HeI{He\,{\sc i}}
\def\HeII{He\,{\sc ii}}

\def\HII{H\,{\sc ii}}
\def\HI{H\,{\sc i}}

\def\OI{O\,{\sc i}}

\def\MgII{Mg\,{\sc ii}}
\def\FeII{Fe\,{\sc ii}}
\def\FeIII{Fe\,{\sc iii}}

\def\SiII{Si\,{\sc ii}}

\def\ie{\hbox{i.e.,}} 
\def\eg{\hbox{e.g.,}}

\def\Kel{\hbox{\rm K}}

\def\Mdot{\hbox{$\dot {M}$}}

\def\Rsun{\hbox{\it R$_\odot$}}

\def\Rstar{\hbox{\it R$_*$}}
\def\Lsun{\hbox{\it L$_\odot$}}
\def\Lstar{\hbox{\it L$_*$}}
\def\Msun{\hbox{\it M$_\odot$}}

\def\Msunyr{\hbox{\it M$_\odot\,$yr$^{-1}$}}

\def\Teff{\hbox{\it T$_{\rm eff}$}}
\def\Vinf{\hbox{$v_\infty$}}
\def\kms{\hbox{km$\,$s$^{-1}$}}

\def\simgr{\mathrel{\hbox{\rlap{\hbox{\lower4pt\hbox{$\sim$}}}\hbox{$>$}}}}

\def\km/s{km~s$^{-1}$}
\def\um{${\mu}$m}

\def\Vinf{\hbox{$V_\infty$}}
\def\HeI{He\,{\sc i}}
\def\HeII{He\,{\sc ii}}
\def\HII{H\,{\sc ii}}

\def\MgII{Mg\,{\sc ii}}
\def\FeII{Fe\,{\sc ii}}

\def\SiII{Si\,{\sc ii}}
\def\Mdot{\.{M}}
\def\FMM362{FMM362}

\defcitealias{naj04}{Paper I}

\shorttitle{Metallicity in the Quintuplet cluster}
\shortauthors{Najarro et al.}

\begin{document}
\title{Metallicity in the Galactic Center: \\ The Quintuplet cluster}

\author{
Francisco Najarro\altaffilmark{1}, Don F. Figer\altaffilmark{2}, 
D. John Hillier\altaffilmark{3},  T. R. Geballe\altaffilmark{4}, Rolf P. Kudritzki\altaffilmark{5}}

\email{najarro@damir.iem.csic.es}

\altaffiltext{1}{Instituto de Estructura de la Materia, CSIC, Serrano 121, 29006 Madrid, Spain }
\altaffiltext{2}{Chester F. Carlson Center for Imaging Science, Rochester Institute of Technology, 54 Lomb Memorial Drive, Rochester, NY 14623}
\altaffiltext{3}{Department of Physics and Astronomy, University of Pittsburgh, 3941 O'Hara Street, Pittsburgh, PA 15260}
\altaffiltext{4}{Gemini Observatory, Hilo, 670 N. A'ohoku Pl., HI 96720}
\altaffiltext{5}{Institute for Astronomy, University of Hawaii, 2680 Woodlawn Drive, Honolulu, HI 96822}

\begin{abstract}

We present a measurement of metallicity in the Galactic center Quintuplet
Cluster made using
quantitative spectral analysis of two Luminous Blue Variables (LBVs).
The analysis employs line-blanketed NLTE
wind/atmosphere models fit to high-resolution near-infrared spectra
containing lines of H, \HeI, \SiII, \MgII, and \FeII.  We are able to
break the H/He ratio vs. mass-loss rate degeneracy found in other LBVs and
to obtain robust estimates of the He content of both objects. Our results
indicate solar iron abundance and roughly twice solar abundance in the
$\alpha$-elements. These results are discussed within the framework of
recent measurements of oxygen and carbon composition in the nearby
Arches Cluster and iron abundances in red giants and supergiants
within the central 30~pc of the Galaxy. The relatively large enrichment of
$\alpha$-elements with respect to iron is consistent with a history of
more nucleosynthesis in high mass stars than the Galactic disk. \end
{abstract}


\section {Introduction}

Elements heavier than hydrogen and helium (``metals'') are primarily
created by nucleosynthesis in stars. Metals are important ingredients in
many astrophysical processes such as radiative cooling, and mass-loss
during star formation and at all stages of
stellar evolution. They also play a fundamental role in stellar evolution
through their influence on stellar opacities, and represent a historical
record of galactic chemical enrichment via stellar winds and supernovae
ejecta.

In the Galaxy metal abundance increases with decreasing galactocentric
radius, as seen in stars and gas
\citep{affler97,rud06,mac99,fuhr98,roll00,sma01,luck06}. Other galaxies show a similar
trend, having highest metal abundances in their nuclei
\citep{urba05,kenni03}.

Previous work \citep{fro99,fel00,car00,ram97,ram99,ram00} on the Galactic
center (GC) has indicated roughly solar stellar metal abundances, whereas
the analyses of interstellar emission lines \citep{shi94,mae00} have
suggested considerably higher abundances. It is not clear why the stellar
and gas-phase measurements should differ so greatly.

The GC contains three dense and massive star clusters that have recently
formed in the inner 50 pc, the Arches, Quintuplet, and Central clusters.
Using quantitative spectral analysis, \citet{naj04} (Paper I) determined
that the WNL stars in the very young (2-2.5~Myr) Arches Cluster have
roughly solar metallicities. Being more evolved ($\sim$4Myr), the
Quintuplet Cluster \citep{gla87,gla90,nag90,oku90,mon94} contains a
variety of massive stars, including WN, WC, WN9/Ofpe, luminous blue
variables (LBVs) and less evolved blue-supergiants
\citep{fig95,fig99a,fig99b}. Two LBVs in it are known, the Pistol Star
\citep{mon94,cot94,fig95,fig98,fig99c} and FMM362 \citep{fig99b,geb00},
each having an infrared spectrum rich in metal lines of \FeII, \SiII, \&\
\MgII.

In this paper, we use quantitative infrared spectroscopy of the two
Quintuplet LBVs to make direct determinations of metallicity in those
stars. We also use the derived $\alpha$-elements vs. Fe ratio to address
the dominance of massive stars on the IMF in this region.

\begin{figure*}
\epsscale{1.05}     
\vspace{-.1cm}
\plotone{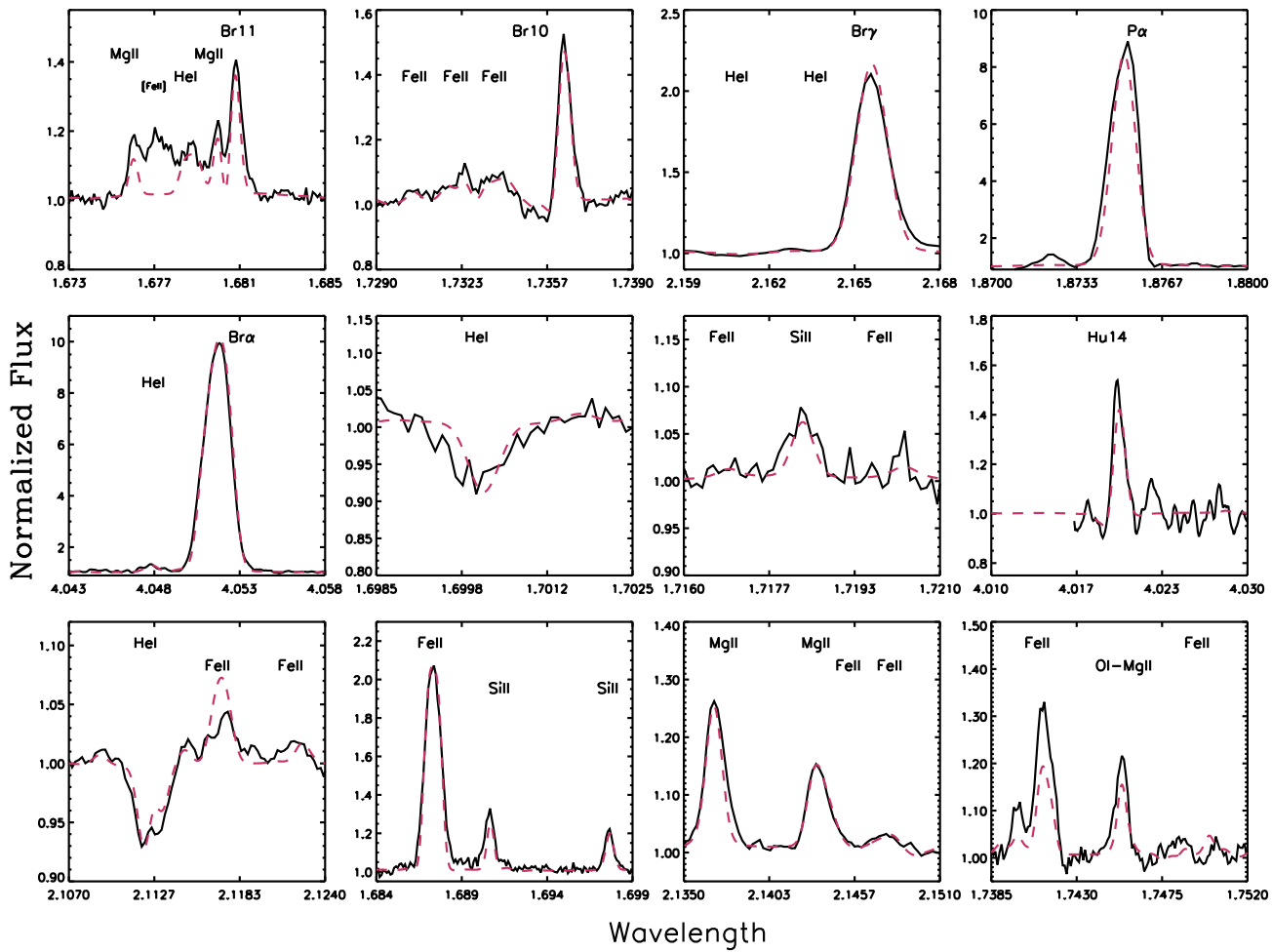}
\caption{\label{fig:pistol}
Model fits ({\it dashed lines}) to the observed infrared diagnostic lines 
({\it solid lines}) of the Pistol Star. The forbidden [\FeII] line at 
1.677~$\mu$m was not included in the models.}

\end{figure*}

\section {Observational Data}

The data were obtained at UKIRT\footnote{The United Kingdom Infrared
Telescope (UKIRT) is operated by the Joint Astronomy Centre on behalf of
the Particle Physics and Astronomy} using CGS4.  The Pistol Star was
observed in April 1996 (P$\alpha$; R$\sim$3000), July 1997 (L;
R$\sim$16000) and April 1998 (H; R$\sim$5000 and K R$\sim$3000). Likewise,
spectra for \FMM362 were obtained in April (L) and May 1999 (H and K),
using CGS4 in medium resolution mode (R$\sim$5,000-6,500).  The slit width
was 0$\farcs$6 for all observations.  We used the photometric measurements
of \citet{fig98} for the Pistol Star, to scale the reduced spectra. For
\FMM362, given its photometric variability, we adopted the average value,
K=7.30, obtained by \citet{gla99} for the epoch closest to our
spectroscopic observations. This value agrees, within the 0.26 standard
deviation derived by \citet{gla99}, with the K=7.50 value adopted by
\citet{geb00} from flux-calibrated spectra. We assume the same extinction
for both objects and adopt the value of A$_K$=3.2 derived by \citet{fig98}
for the Pistol Star. The reader is referred to these papers for a detailed
discussion on the reduction of the observed spectra and photometry.

\begin{figure*}
\epsscale{1.00}
\vspace{-.1cm}
\plotone{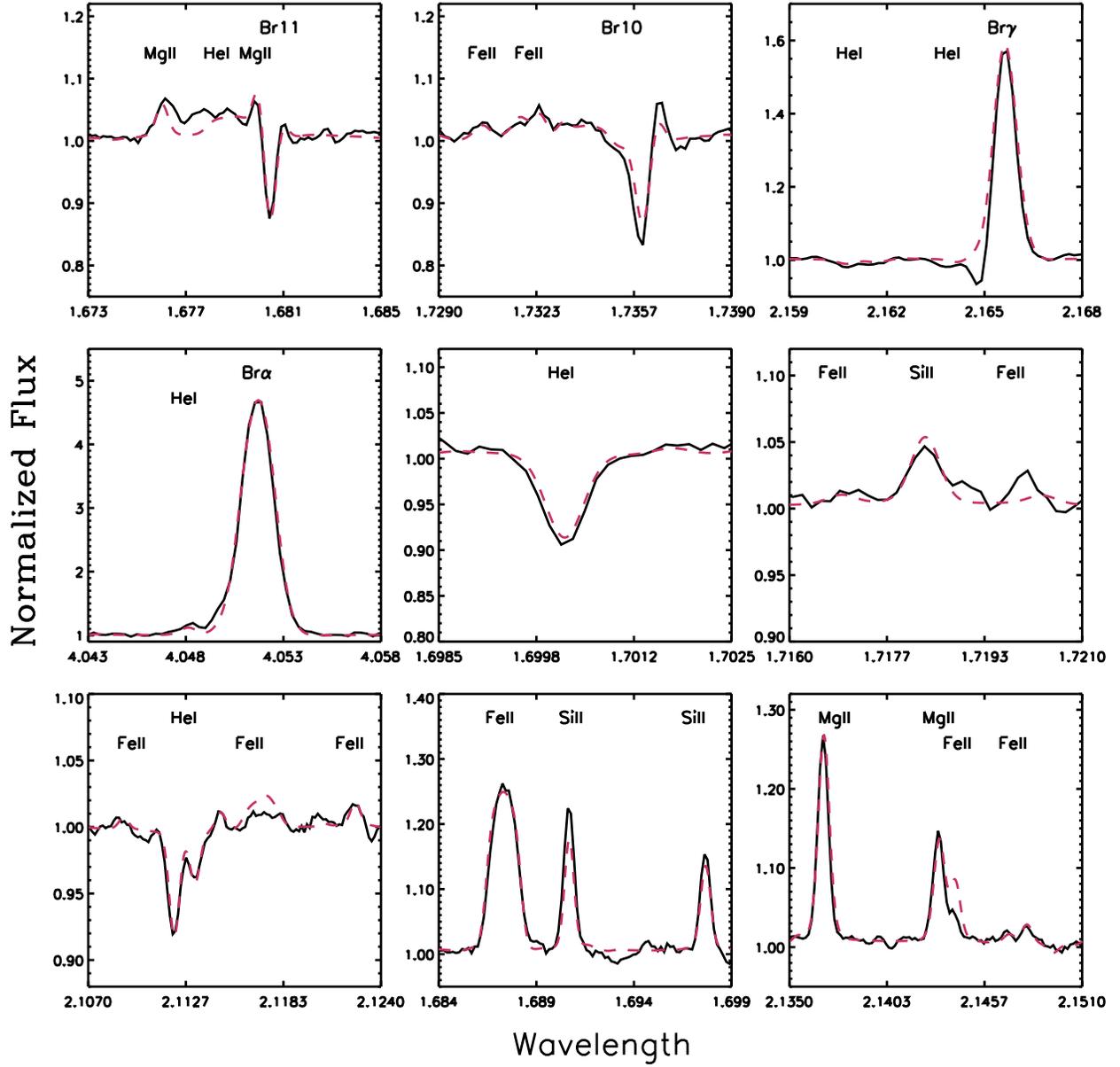}
\caption{\label{fig:362}
Model fits ({\it dashed lines}) to the observed infrared diagnostic lines 
({\it solid lines}) of \FMM362.}
\end{figure*}

\section{Models}

To model the LBVs and estimate their physical parameters, we have used CMFGEN,
the iterative, non-LTE line blanketing method presented by \citet{hil98} which
solves the radiative transfer equation in the co-moving frame and in
spherical geometry for the expanding atmospheres of early-type stars. The
model is prescribed by the stellar radius, \Rstar, the stellar luminosity,
\Lstar, the mass-loss rate, \Mdot, the velocity field, $v(r)$ (defined by
\Vinf\ and $\beta$), the volume filling factor characterizing the clumping
of the stellar wind, {\it f(r)} (see Sec.~\ref{sub-clu}), and elemental
abundances. \citet{hil98,hil99} present a detailed discussion of the code.
For the present analysis, we have assumed the atmosphere to be composed of
H, He, C, N, O, Mg, Si, S, Fe and Ni. 
Given the parameter domain the LBVs are located, the $\tau=2/3$ radius 
is located close or above the sound speed, and therefore the assumed
hydrostatic structure plays no role. Thus, no spectroscopic information
about the mass of the object can be obtained.
The atomic data sources are described in detail \citet{hil01}.
Here we focus on the model atoms used for our abundance determinations \FeII,
\MgII\ and \SiII. Using the superlevel formalism 
\citep[NS/NF, number of superlevels vs. number of levels in the full atom, e.g.][]{hil98} we
we chose 233/709 (up to 109800~cm$^{-1}$) for \FeII,
37/50 (up to 119400~cm$^{-1}$) for \MgII\ and 35/72 
(up to 125000~cm$^{-1}$) for \SiII.
The choice of the appropiate packing has been  extensively tested in
\citet{naj01}. We will revise the importance of this issue for the 
case of Mg in \ref{sub-meta}.

Observational constraints are
provided by the H, K and L-band spectra of the stars and the dereddened K
magnitudes from \citet{fig98}, \citet{geb00} and \citet{gla99}.
As in \citetalias{naj04},
a distance of 8~\kpc\ has been assumed. The validity of our technique has
been demonstrated in \citet{naj99} and \citet{naj01} by calibrating our
method against stars with similar spectral type such as P~Cygni and
HDE~316285 for which not only infrared but also optical and UV spectra
are available.

\begin{deluxetable}{lll}
\tablecaption{Derived stellar parameters}
\tablehead{
\colhead{{Parameter}} &
\colhead{{Pistol}} &
\colhead{{\FMM362}}} 
\startdata
\Lstar ($10^{6}\Lsun$)   & 1.60 & 1.77   \\
\R23 (\Rsun)             & 306  & 350    \\
\Teff  ($10^{4}$K)       & 1.18 & 1.13   \\
H/He                     & 1.5 & 2.8   \\
Fe/Fe$_\odot$            & 1.1 (0.78) & 1.1 (0.78)   \\
Mg/Mg$_\odot$            & 2.2   & 1.5   \\
Si/Si$_\odot$            & 1.8   & 2.1  \\
\Mdot ($10^{-5}\Msunyr$) & 2.1 & 1.2   \\
$\beta$                  & 3.0  & 1.3   \\
\Vinf (\kms)             & 105  & 170   \\
D$_{mom}$=log(\Mdot\Vinf(R/\Rsun)$^{1/2}$) & 29.39 & 29.38   \\
 M$_{Edd}$ (\Msun)   & 22.5 & 30.5  \\
 {{CL$_1$}}    & 0.08 & 0.08   \\
 {{CL$_2$}}    & 2.5  & 2.00   \\
 {{CL$_3$}}    & 2.00 & --   \\
\enddata
\tablecomments{$\beta$ is the exponent describing the velocity field,
D$_{mom}$ is the modified wind momentum \citep{kud00} and 
clumping parameter, the CL are defined in Eq.~\ref{eq:clump}. M$_{Edd}$ are
the Eddington masses.
H/He is the ratio by number, and other abundances are relative to solar 
after \citet{gre93}. The Fe abundance relative the value in \citet{and89} 
is given in parentheses (see text).}
\end{deluxetable}

\section{Results}
\label{sec-results}

Table~1 gives the derived stellar parameters for both LBVs and
Figs.~\ref{fig:pistol} and \ref{fig:362} show model fits to the relevant
lines in the stars. Theoretical spectra have been convolved with the
instrumental resolution. We note that given the large number of parameters
involved in the analysis it is
unafordable to perform a full systematic error analysis in the whole
parameter domain. We rather proceed by estimating the range of values for the
main stellar parameters which provide acceptable fits to the observed spectra.
Once those ranges are set, we derive the corresponding abundances and their
errors.
From Table~1, it can be seen that the Pistol Star
and \FMM362 have very similar properties, with the exceptions of the
Pistol Star's significantly higher wind density (evidenced by the its
stronger spectral lines) and its higher He content. The latter may denote
a slightly advanced evolutionary stage for the Pistol Star (see below).  
Given the general resemblance of the spectra of the objects, we discuss
them together.

\subsection{Main Diagnostic Lines and  Stellar Properties}
\label{sub-prop}

Several spectral diagnostics constrain our estimates of the stellar
temperature, and thus the ionization structure, in particular the \HeI\
(5-4) components near 4.05~\um.
If helium is predominantly singly ionized, even for the most
favorable case with (minimum) cosmic helium abundance,
the observed ratio of H to  \HeI\ lines exceeds the expected values by 
large factors.
This indicates that \HeII\ must recombine to \HeI\ very close
to the photosphere, implying an upper limit of around 13,000~K for the
temperatures of these objects. We find a lower limit of 10,000~K for the
temperatures, as lower values would require non-detection of the \HeI\
components.  Also the strengths of the \hetrit\ and \hedubi\ lines are
very sensitive to temperature, so that they appear in emission above
12,500~K and vanish below 10,500~K. These lower limits on the effective
temperature are also consistent with the non-detections of the \SiII\
3s$^2$3p$^2$S$_{1/2}$-3s$^2$4p$^2$P$_{3/2}$~2.180~\um\ and
3s$^2$3p$^2$S$_{1/2}$-3s$^2$4p$^2$P$_{1/2}$~2.209~\um\ intercombination
lines, as they are expected to appear strongly in absorption as soon as
the temperature drops below 10,000~K. Hence, $\Delta$T$\pm1500$~K are
conservative estimates of uncertainties for the temperatures of
the Pistol Star and \FMM362 given in Table~1.

\begin{figure*}
\epsscale{1.0}
\vspace{-.1cm}
\plotone{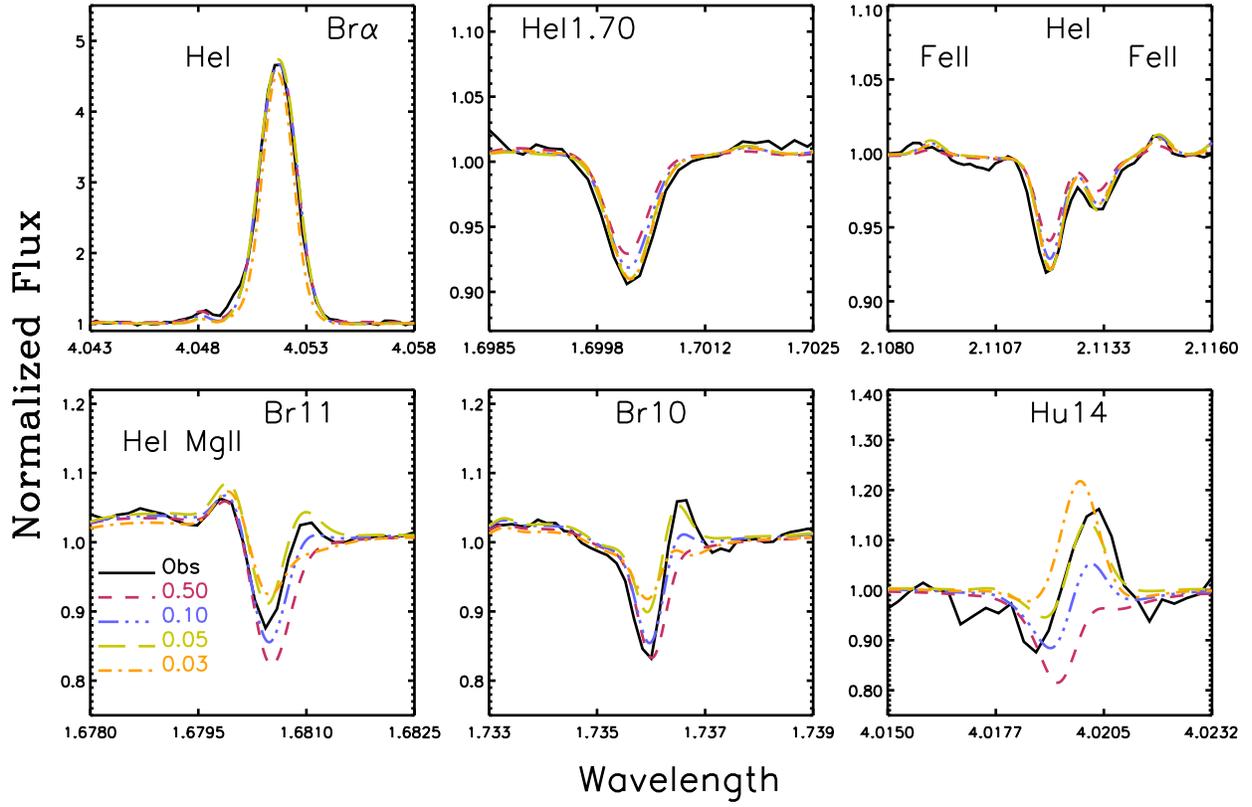}
\vspace{-.1cm}
\caption{\label{fig:l362_clump}
Model spectra showing the effect of clumping parameter CL$_1$
(see Eq.\ref{eq:clump}) on diagnostic line profiles of \FMM362. Values of 
CL$_1$ are listed in the bottom left panel.} 
\end{figure*}

To estimate the terminal velocities, we make use of the \FeII]
(semi-forbidden \FeII) z$^{4}$F$_{9/2}$-c$^{4}$F$_{9/2}$ 1.688~\um\ line
\citep{geb00,fig98} that forms in the outer wind and has a weak oscillator
strength (gf$\sim$10$^{-5}$). This is because non-negligible continuum 
opacity effects at 4~\um\ may provide only lower limits if Br~$\alpha$ is 
used. The larger \Vinf\ derived for \FMM362\ can be clearly inferred from 
the width of this \FeII] line and  the obvious overlap at 
\bap\ between the \HI\ and \HeI\ components (see Fig.~\ref{fig:362}). For 
the Pistol Star (Fig.~\ref{fig:pistol}) the \bap\ components are fairly 
well separated.

Our analysis of wind density (\Mdot, $\beta$) gives values of $\beta$ that
agree with those inferred in the literature for other LBVs \citep{naj01}
and B-Supergiants \citep[e.g][]{cro06}. The values are fairly well
constrained by the shapes of the hydrogen lines, especially those of \bap\
and \bgam\, which are inconsistent with the same value of $\beta$
for both objects.

Although the wind density derived for the Pistol Star is much higher than
for \FMM362, the modified wind momenta D$_{mom}$ = log (\Mdot \Vinf
$\sqrt{R/\Rsun}$) \citep{kud00} of the two LBVs are nearly identical
(Table~1). This result is qualitatively consistent with the wind momentum
- luminosity relation \citep{kud00} which predicts the same modified
momenta for objects with the same stellar type and luminosity. Note that
we have used clumping-corrected values of \Mdot\ to compute the modified
momenta. If unclumped values were assumed, D$_{mom}$ would be closer to
30.0. Interestingly, the latter agrees very well with the averaged
modified momentum of AG~Car at maximum (\ie\ at similar \Teff\ to our
objects) obtained using the values of \Mdot, \Vinf\ and \Rstar\ derived by
\citet{sta01} from fits to \hap. Those authors obtained
log(\Mdot)$\sim$-4.1 for about this temperature, while our unclumped
values are log(\Mdot)=-4.1 and -4.4 for the Pistol Star and \FMM362\,
respectively. Further, using radiation-driven wind models for LBVs,
\citet{vin02} were able to predict the \citet{sta01} \Mdot\ value for
AG~Car assuming \Vinf/\Vesc$\sim$1.3 and a current stellar mass of
35\Msun. The same value of \Vinf/\Vesc\ for the LBVs would imply current
stellar masses of 27.5\Msun\ for the Pistol Star and 46\Msun\ for \FMM362.
Although these masses should be regarded with caution, they are consistent
with the Pistol Star being more evolved than \FMM362 (as inferred from
their He/H ratios) and hence having lost more mass during its evolution,
as indicated by the presence of a nebula around it.

Compared to the results obtained in \citet{fig98} by means of
non-blanketed models, the new blanketed models provide a significant
improvement in our knowledge of the physical properties of these two
stars. The degeneracy of the ``high'' and ``low'' luminosity (\Teff)
solutions for the Pistol Star presented by \citet{fig98} is broken by the
\SiII, \MgII\ and \FeII\ lines, which are clearly more consistent with the
``low'' solution. We derived for this star a luminosity of
$\sim$1.6($10^6$)~\Lsun, an effective temperature of $\sim$11,800~\Kel,
and an initial mass of 100~\Msun. The stellar luminosity is reduced by a
factor of two compared with the previous estimate, illustrating the
importance of the new generation of line-blanketed models.  Below, we
discuss in detail the role of two additional stellar properties that are
derived using the new models, wind clumping and elemental abundances.

\subsection{Clumping}
\label{sub-clu}

Clumping is normally invoked in stellar winds to explain inconsistencies
arising between $\rho$ (density) and $\rho^2$ diagnostics. For a given
mass-loss, clumping causes an enhancement of $\rho^2$ processes while
leaving unaltered those which depend linearly on $\rho$.  Further, if
mass-loss rate and clumping are scaled without changing the
\Mdot$/f^{0.5}$ ratio, the $\rho$-dependent diagnostics vary while
the recombination lines profiles ($\propto \rho^2$) remain basically
unaltered.

To investigate the clumping we introduce the following clumping law:

\begin{equation}
\label{eq:clump}
  f = CL_1 + ( 1 -CL_1 )  e^{\frac{-V}{CL_2}} + ( CL_4 - CL_1 ) 
e^{\frac{(V-V_{\infty})}{CL_3}}
\end{equation}

\noindent where CL$_1$ and CL$_4$ are volume filling factors and CL$_2$ and CL$_3$
are velocity terms defining locations in the stellar wind where the
clumping structure changes. CL$_1$ sets the maximum degree of clumping
reached in the stellar wind (provided CL$_4$$>$CL$_1$) while CL$_2$
determines the velocity of the onset of clumping. CL$_3$ and CL$_4$
control the clumping structure in the outer wind. Hence, when the wind
velocity approaches \Vinf, so that (V-\Vinf)$\leq$CL$_3$, clumping starts
to migrate from CL$_1$ towards CL$_4$. If CL$_4$ is set to unity, the wind
will be unclumped in the outermost region.  Such behavior was already
suggested by \citet{nug98} and was utilized by \citet{fig02} and
\citet{naj04} for the analysis of the WNL stars in the Arches Cluster.
Recently, \citet{pul06} also have found similar behavior from \hap\ and
radio studies of OB stars with dense winds. Furthermore, our clumping
parametrization is consistent with results from hydrodynamical
calculations by \citet{run02}. From Eq.~\ref{eq:clump} we note that if
CL$_3$, and therefore CL$_4$, is not considered (CL$_3\rightarrow$0), we
recover the simpler variation proposed by \citet{hil99}. To avoid entering free
parameters heaven we set CL$_4$=1 in all of our investigations, aiming to
get an appropriate amount of leverage on the amount of non-constant
clumping in the outer wind regions.

\begin{figure*}
\epsscale{1.15}    
\vspace{-.1cm}
\plottwo{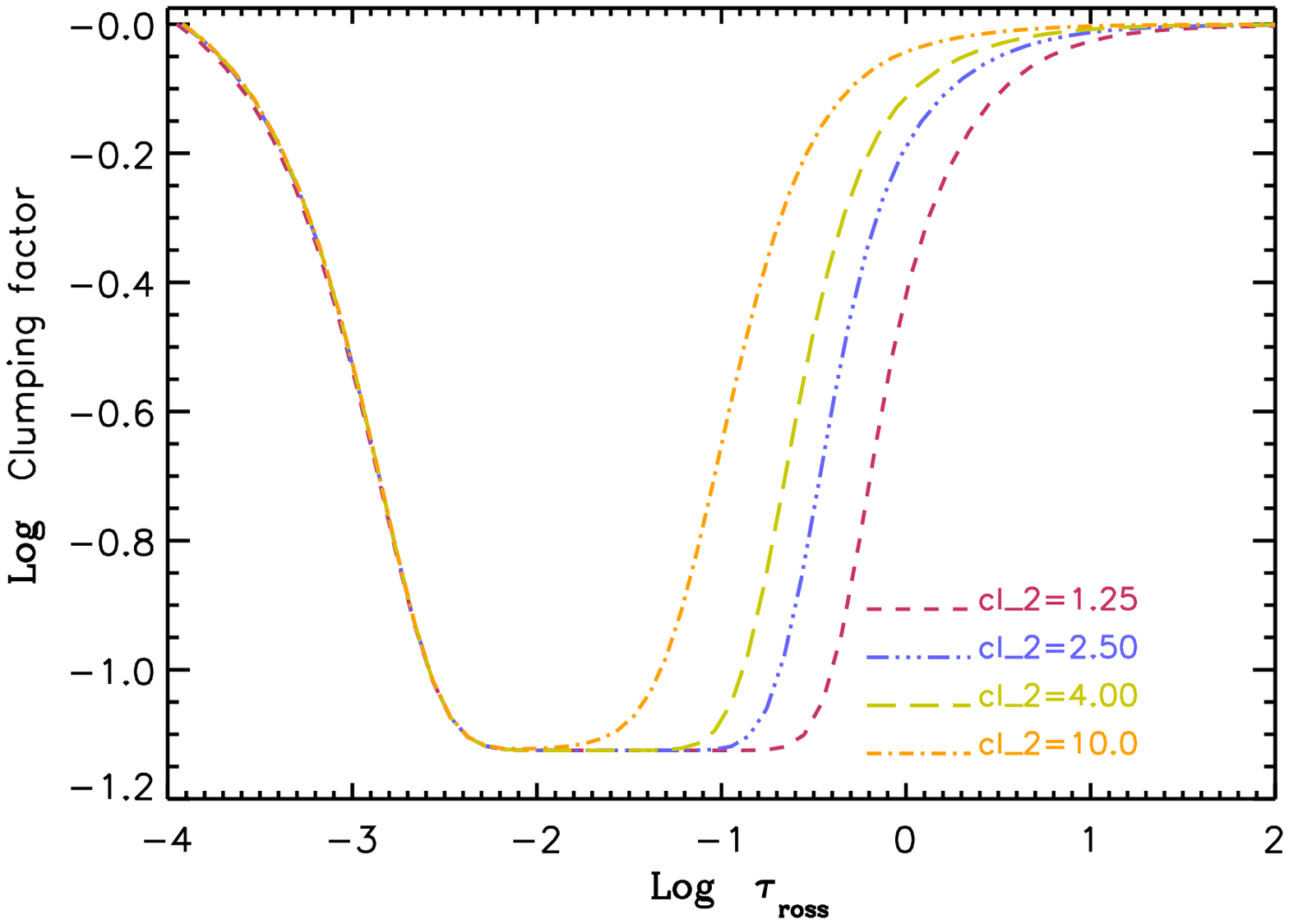}{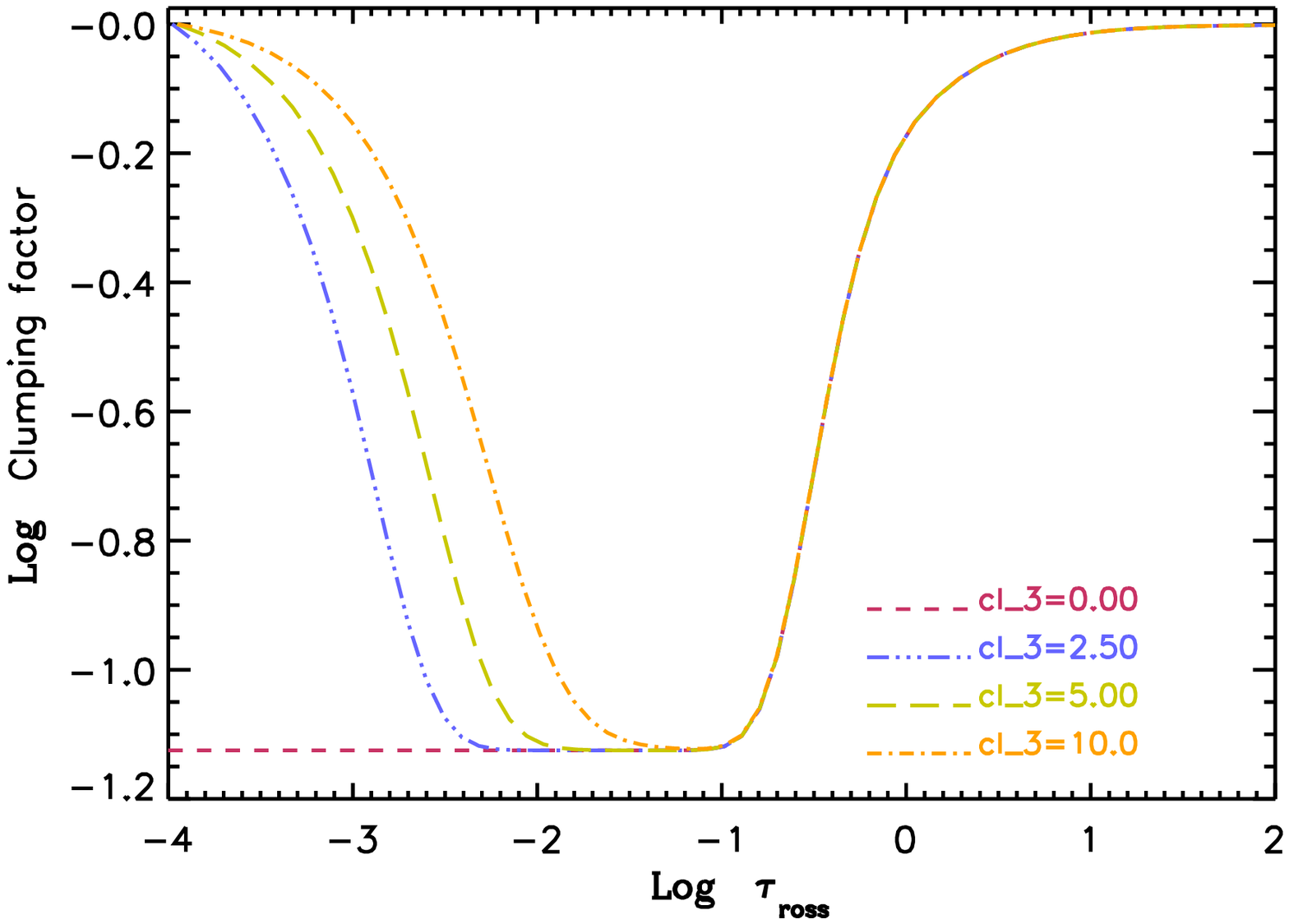}
\vspace{-.1cm}
\plottwo{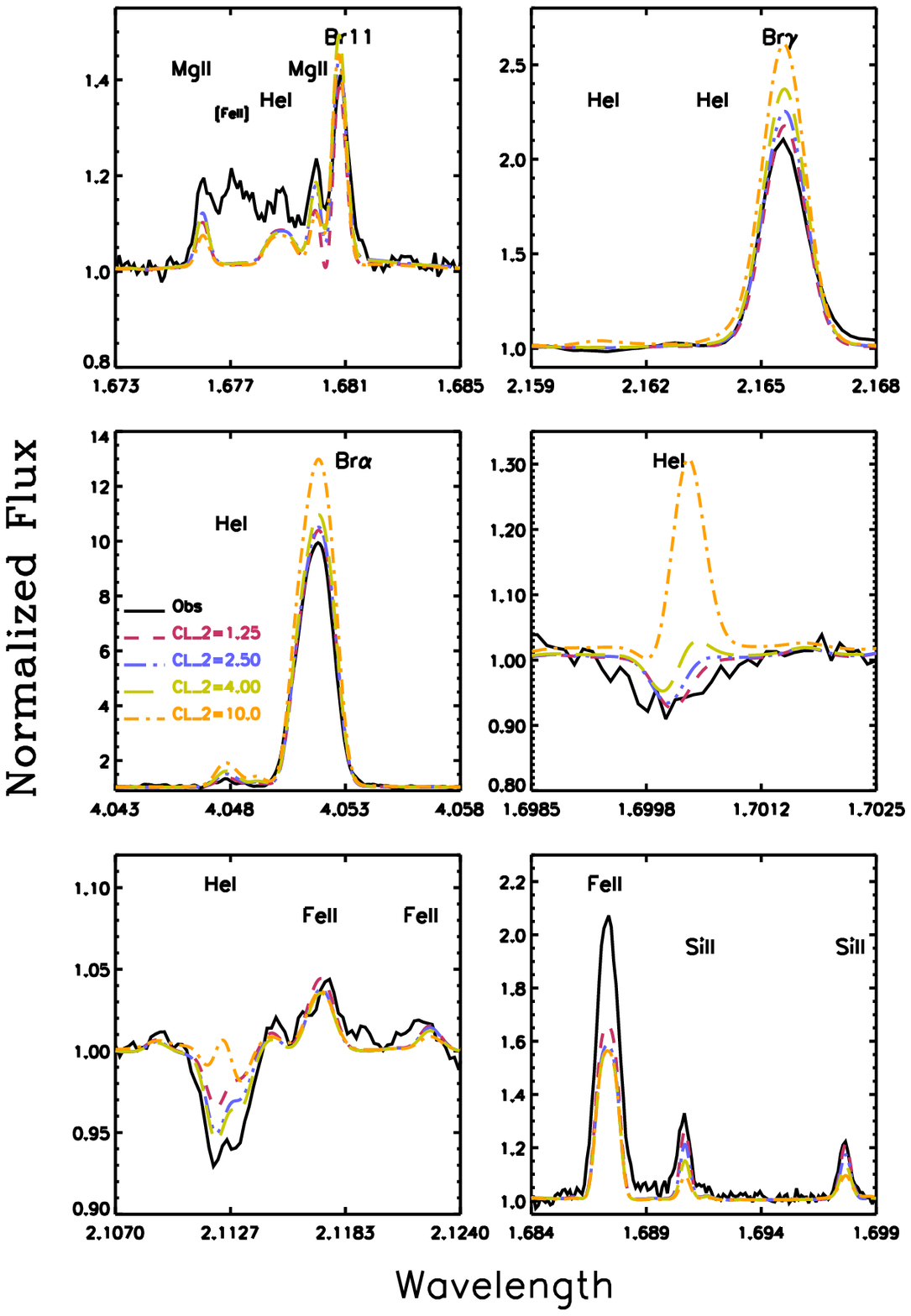}{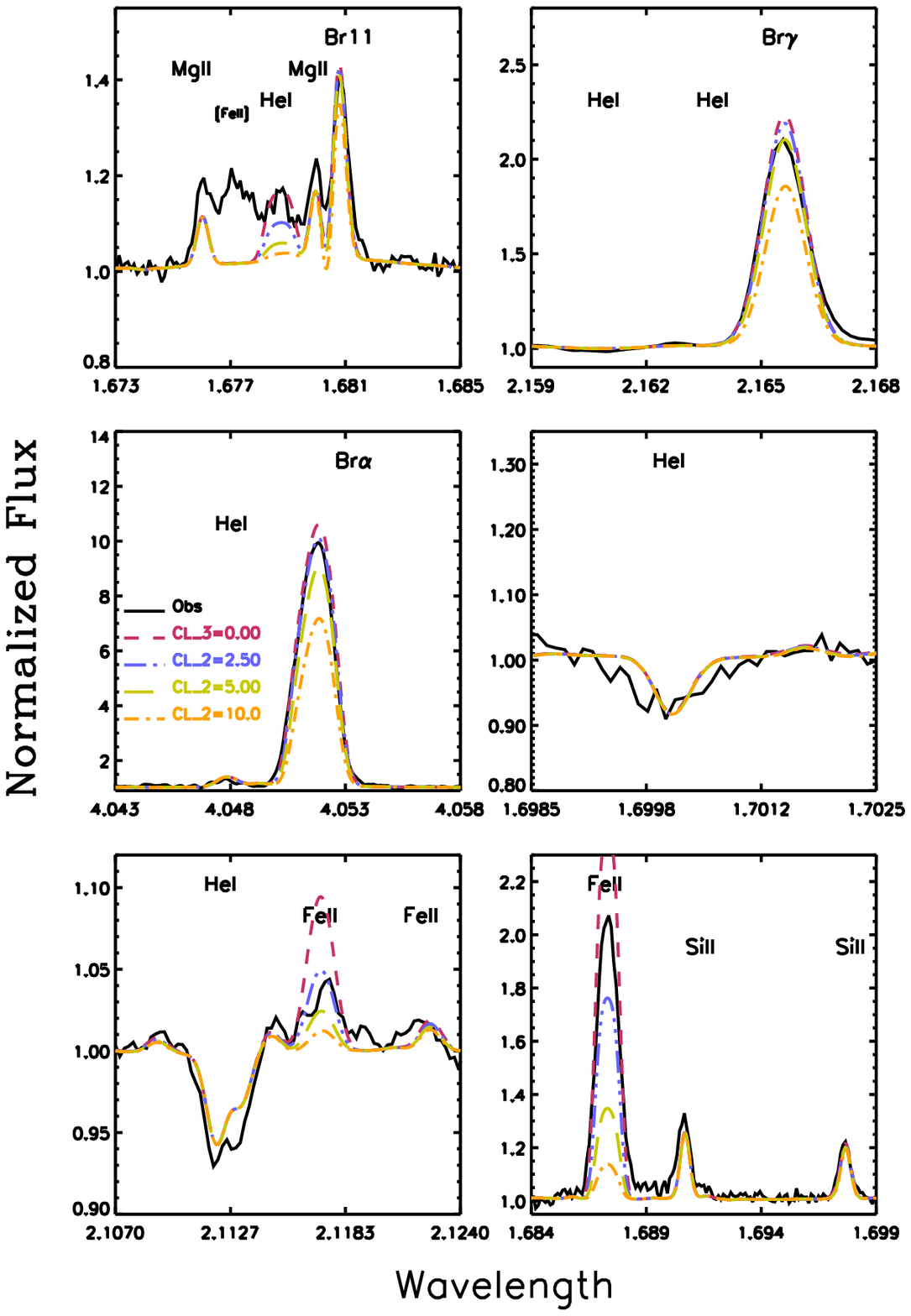}
\caption{\label{fig:pistol_clump}
Influence of adopted clumping structure
(see Eq.~\ref{eq:clump})
in profiles.  Run of clumping
and line profile sensitivities for different values of CL$_2$
({\it{left panel}}) and CL$_3$ ({\it{right panel}}), compared to observed 
profiles in the Pistol Star. See discussion in text.}
\end{figure*}

{\bf{CL$_1$: Estimating the wind clumpiness.}} Figure~\ref{fig:l362_clump}
illustrates the sensitivity to CL$_1$ of the main diagnostic lines
utilized to obtain the degree of clumping in the stellar winds of \FMM362.
For each value of CL$_1$ displayed in Fig.~\ref{fig:l362_clump} the
mass-loss rate in the model was scaled while keeping \Mdot$/f^{0.5}$
constant as described previously. Although there are some lines that
follow this scaling quite well (\bap, and also \bgam\ and some metal lines
not displayed in the figure), the \HeI\ lines and weak \HI\ lines react
quite sensitively to the absolute degree of wind clumping. It can be seen
that the \HeI\ lines only provide an upper limit to CL$_1$ (a bit lower
than 0.1) and do not react to lower values, but a unique value for
CL$_{1}$ can be selected by some \HI\ lines. The Hu$_{14}$ (\HI(14-6))
line displays the highest sensitivity to clumping.  Unfortunately, the
wavelength interval surrounding this line was not observed with
sufficiently high S/N in \FMM362\ and there is a large uncertainty in the
continuum value, which is critical for estimating CL$_{1}$. We could make
full use of the Hu$_{14}$ line to determine the clumping only for the
Pistol Star, where this line is relatively stronger in emission (see
Fig.~\ref{fig:pistol}). Nevertheless, Fig.~\ref{fig:l362_clump} shows that
CL$_1$ lies between 0.1 and 0.05. It must be stressed that only with a
well determined clumping may we address the He/H abundance issue (see
Sec.~\ref{sub-hhe}).

{\bf{CL$_2$ \&\ CL$_3$: Mapping the clumping structure.}} The upper panels
of Fig.~\ref{fig:pistol_clump} illustrate the behavior of the clumping
structure for different sets of CL$_2$ and CL$_3$ values, while the lower
panels display the influence of such behavior on diagnostic lines in the
spectrum of the Pistol Star. It is evident that for some spectral lines
\eg~\hedubt, the behavior of the profiles with clumping is far from being
monotonic. Furthermore, not only do lines of different ions react
differently to clumping, but also lines within the same ion, \eg~\HI,
behave differently. For example, \FeII\ does not respond in the same way
to changes in CL$_2$ and CL$_3$. Figure~\ref{fig:pistol_clump} shows the
great potential of the different IR lines to constrain the clumped
structure of the stellar wind and demands the following detailed
discussion.

The general impact of clumping on line profiles that was described at the
beginning of this section will occur provided the ionization equilibrium
is on the ``safe'' side. We consider the ``safe'' region to be where the
population of the next ionization stage clearly dominates over the one the
line belongs to (\ie\ \HII$\gg$\HI\ for the hydrogen lines). Noting,
however, that ionization depends linearly on density whereas recombination
is proportional to $\rho^2$, a ``changing'' ionization situation may
occur, where two adjacent ionization stages have similar populations. In
such a case clumping, which enhances recombination, will cause a net
reduction of the mean ionization. This will result in weaker lines.

Finally, in the infrared, via bound-free and free-free processes 
$\propto \rho^2$), not only the lines but also the continuum will depend 
on clumping, resulting in high sensitivity of the continuum-rectified
line profiles to CL$_1$ and CL$_2$. 

One may, therefore, distinguish between lines formed on the ``safe''
region and those arising from the ``changing'' region.  Within the
parameter domain of the two LBVs studied here we find that \HI, \SiII, and
\MgII\ lines and also the \FeII\ photospheric lines are formed in
``safe'' regions, while \HeI\ and \FeII] lines arise from
``changing'' regions.

Increasing the clumping (decreasing the CL$_1$ value), or alternatively
decreasing the velocity at which clumping sets in (decreasing CL$_2$)
results in stronger \SiII\ and \MgII\ lines, as shown in
Fig.~\ref{fig:pistol_clump}-left.  The strong \HI\ lines are formed
further out than the \SiII\ and \MgII\ lines and their strengths should in
principle show no sensitivity to CL$_2$. However, the continuum is clearly
affected by clumping. Thus the stronger the clumping (lower CL$_2$), the
stronger the continuum and the weaker the resulting line-to-continuum
ratio, as clearly shown by \bgam\ and \bap\ in
Fig.~\ref{fig:pistol_clump}-left.  On the other hand, weaker \HI\ lines
such as Hu$_{14}$ or B$_{10}$ and B$_{11}$ form much closer to the
photosphere and tend to brighten with increasing clumping. The weak \FeII\
lines formed close to the photosphere react in basically the same way as
the continuum and thus the normalized spectra of them show no changes with
changing clumping. Their near independency allows these lines to be used
as Fe abundance indicators (see below).  The \FeII] lines, formed even
beyond the \HI\ lines, are affected by two competing processes. On one
hand, increasing the extent of the clumped region (decreasing CL$_2$)
results in a reduction of the \FeIII/\FeII\ ratio in the wind. Since
\FeIII\ remains the dominant ionization stage in the \FeII] line formation
zone, this change will cause a slight increase in the strengths of the
\FeII] lines. On the other hand, as the continuum increases with
increasing clumping, the line-to-continuum ratio decreases.  Thus the two
processes counter-balance (see Fig.~\ref{fig:pistol_clump}-left). Finally,
the \HeI\ lines, which form close to the photosphere, show weak continuum
dependences, but high sensitivities to ionization/recombination. Thus,
starting with the model with the highest CL$_2$ values, the \HeI\ lines
are not affected by clumping, but the stellar parameters produce strong
ionization in the inner parts resulting in overly strong emission
(\hetrit) and line filling (\hedubt). However, as clumping is enhanced in
the line formation zone, recombination starts to dominate over ionization
and the \HeI\ line emission weakens, the lines are no longer as filled,
and start to appear in absorption.
  
Regarding clumping in the outer parts of the wind, it can be seen on the
right side of Fig.~\ref{fig:pistol_clump} that only the strong \HI\ and
\FeII] lines react to CL$_3$.  Note that \bap\, which forms further out than
\bgam\, is more sensitive to clumping and the observed ratio of the two
profiles may be used to determine CL$_3$. For winds of significantly lower
density, these lines will form further in and show little or no dependence
on CL$_3$ (\eg~\FMM362). The \FeII] lines are more sensitive to CL$_3$,
primarily due to the coupling of the Fe ionization structure with that of
hydrogen thru charge-exchange reactions in the outer wind zones where \HII\
starts to recombine. Due to the difference in H and Fe abundances, a small
and hardly noticeable change in the ionization of hydrogen will be amplified
in the \FeIII/\FeII\ ratio, resulting in a large change in iron
recombination. Thus, decreasing CL$_3$ dramatically enhances the \FeII]
lines, as shown in on the right side of Fig.~\ref{fig:pistol_clump}. The
\HI\ lines behave similarly to the way they do in ``safe'' regions.
  
For the Pistol Star the best fits to the various line profiles suggest a
decreasing clumping factor that becomes unity in the outer wind, whereas for
the less dense wind of \FMM362, the line profiles are best matched by a
constant clumping factor.  From the clumping estimates, we find
uncertainties of 0.10~dex in the mass-loss rates.  The value of CL$_1=0.08$
obtained for each star is somewhat low compared to the values derived for
other LBVs such P~Cygni \citep[CL$_1$=0.5,][]{naj01} or AG~Car
\citep[CL$_1$=0.25,][]{gro06} and is more consistent with those derived for
WR stars \citep[CL$_1$$\sim$0.1,][]{her01,naj04}. We note, however, that the
LBVs have higher He abundances than P~Cygni, pointing to a more evolved
status, closer to the WR phase.

\begin{figure}
\epsscale{1.00}    
\vspace{-.1cm}
\plotone{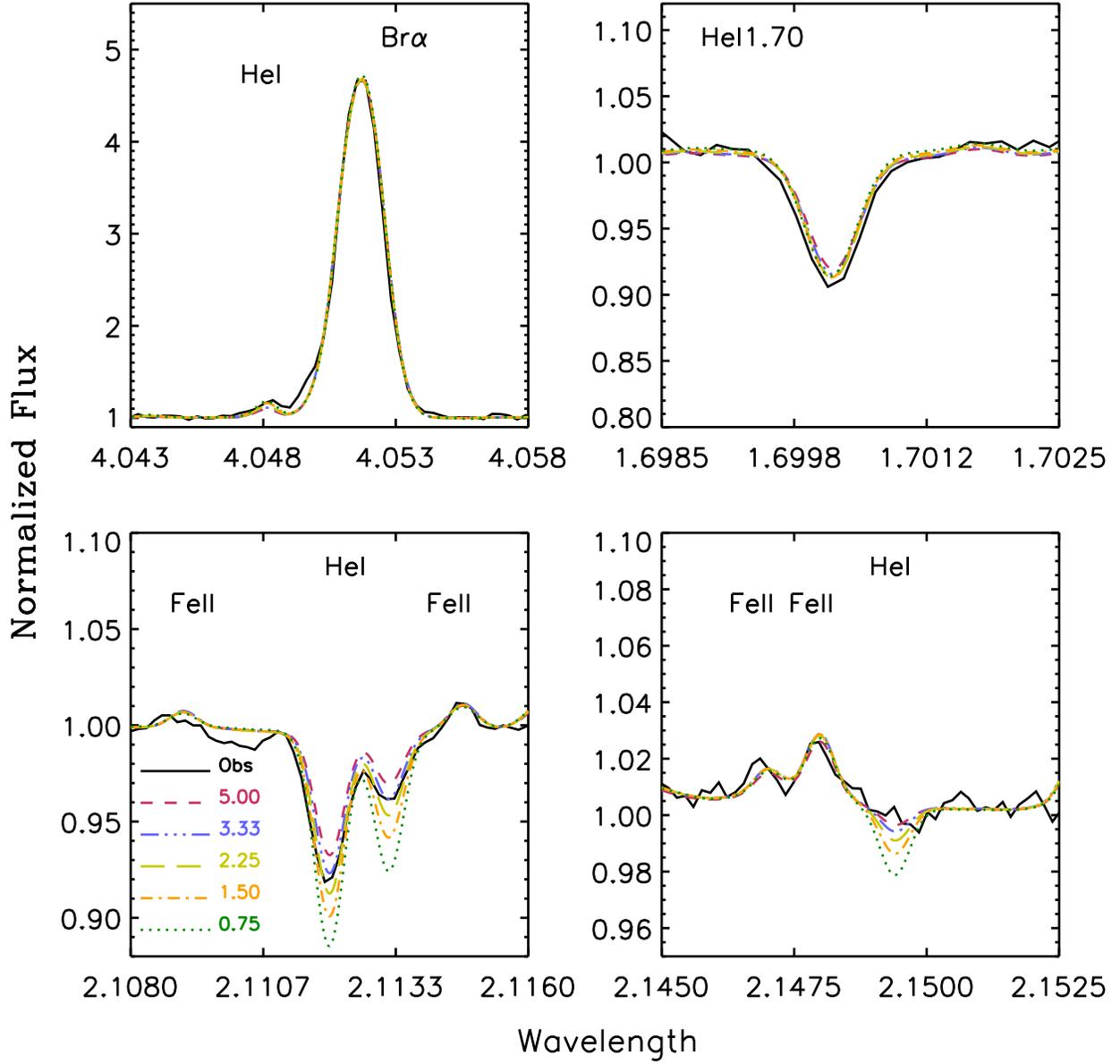}
\caption{\label{fig:hei362}
Breakdown of the H/He degeneracy in \FMM362. Models with H/He ratios
ranging from 5.0 to 0.75 provide identical \HI\ and strong \HeI\ line
profiles, but weaker \HeI\ profiles can be used to determine the He
abundance (see text)}
\end{figure}

\subsection{H/He ratio. Breaking the degeneracy}
\label{sub-hhe} 
 
\citet{hil98b} showed that for HDE~316285, an LBV-like star with similar
\Teff\ and slightly higher wind density than the Pistol Star, a degeneracy
exists between the H/He ratio and the mass-loss rate.  In principle, fits of
virtually equal quality could be obtained with H/He ratios varying from 0.05
to 10 by scaling the mass-loss rate. Such a degeneracy, if present in the
Quintuplet LBVs, would imply that if the H/He ratio falls below a certain
value (H/He$\leq$2), the resulting metal abundances could be scaled down to
obtain the same line strengths. Hence one could obtain only an upper limit
on the metal abundances. breaking this degeneracy is crucial to
understanding the evolutionary status of these objects.
 
Because of the lower wind density of the Quintuplet LBVs and the
sensitivities of some of the infrared lines to the stellar parameters, we
are able to break the H/He degeneracy and obtain robust estimates of their
He content. Due to the high degree of clumping found in both objects, the
$\tau=2/3$ radius, where \Teff\ is defined, is reached at considerably lower
velocities than for classical LBVs. Thus, wind speeds roughly between half
and one-third of the sound speed are found in the Pistol Star and \FMM362\
while classical LBVs have wind speeds well above the speed of sound
\citep{hil98b,naj01}.  This enables quasi-photospheric absorption lines to
form. The  \hedubt\ line is the key in breaking the degeneracy. This is
shown for \FMM362\ in Fig.~\ref{fig:hei362}, which contains model spectra
computed for H/He ratios ranging from H/He=5.0 to 0.75, with mass-loss rates
and metal abundances scaled and the other stellar parameters fine-tuned to
reproduce the observed profiles of other lines.  The figure shows that while
identical \HI\ and \HeI\ (\hetrit) line profiles (also for the rest of
hydrogen and metal lines) are obtained for all H/He ratios considered, the
absorption {\it depths} of the \hedubt\ and \HeI~2.15~\my\ lines react
sensitively to the He abundance.  Both lines show that the best H/He value
must lie between 3.33 and 2.25, and we find a most likely value of 2.8 (see
Table~1). Similar behavior was found for the Pistol Star, where we obtain
H/He=1.5.

\subsection{Metal Abundances}
\label{sub-meta}

For the purpose of discussing metal abundances (see Table~2), we adopt the solar
composition of \citet{gre93}. Although their abundances have been recently
revised \citep{asp05,alle08} (but see also \citet{pin06}), they are the ones used
by \citet{igl96} to compute stellar interior opacities and adopted in the
most recent evolutionary models for massive stars with rotation from the
Geneva group \citep{mey03,mey05}, and the Padova tracks used for cooler,
less massive stars \citep{gir00,sal00}. Previously published evolutionary
tracks for massive stars \citep{scha92,mey94} used opacity tables calculated
with solar composition from \citet{and89}, which differ significantly from
\citet{gre93} only in Fe (A(Fe/H)=7.67 vs 7.50 in
\citet{gre93})\footnote{A(X/Y)=log[n(X)/n(Y)]+12} and very slightly in the
CNO ratios (A(C/H)=8.56, A(N/H)=8.05, A(O/H)=8.93 in \citet{and89} vs
A(C/H)=8.55, A(N/H)=7.97, A(O/H)=8.87 in \citet{gre93}). Nevertheless, we
have also listed in Table~1, in parentheses, the measured abundances with respect to
the solar Fe values from \citet{and89}. Si and Mg are the same in all
evolutionary models, and have been only slightly revised downward
($\sim$0.05~dex) by \citet{asp05}. Thus, the reader should note that current
discussions found in the literature on the
derived $\alpha$-elements vs. Fe ratio may depend critically
on the assumed Fe solar abundance.

\begin{figure*}
\epsscale{1.0}
\vspace{-.1cm}
\plottwo{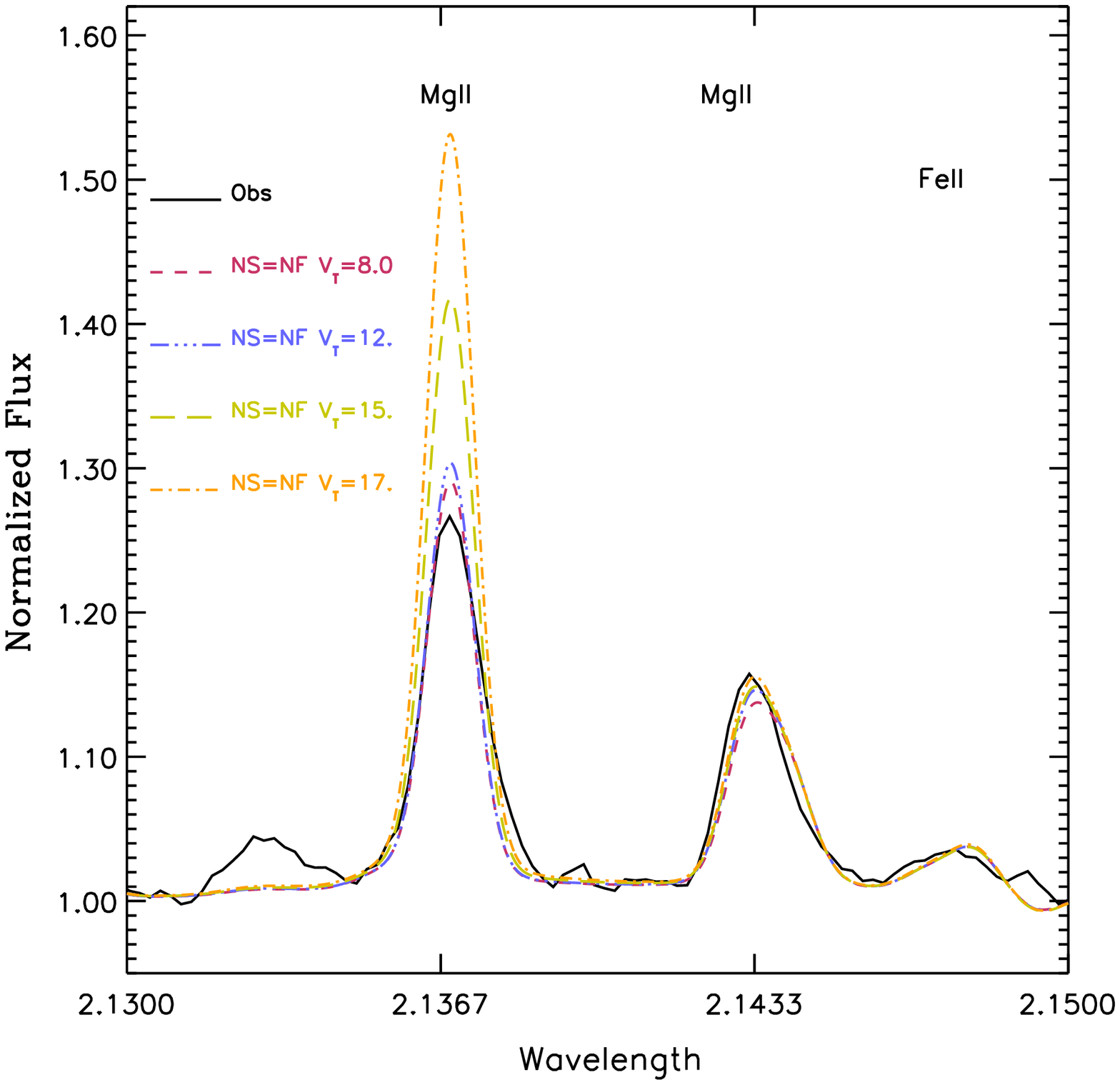}{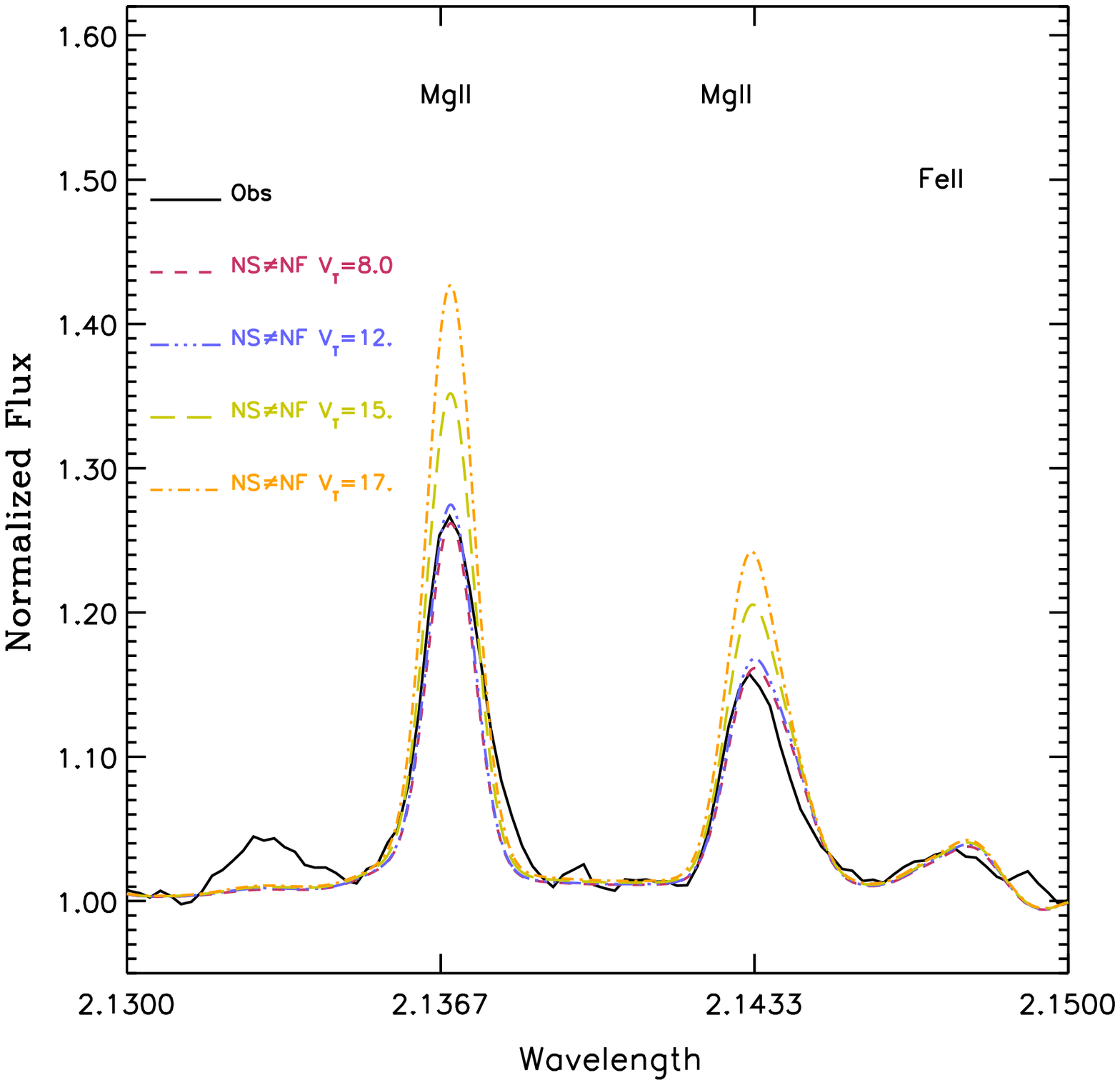}
\caption{\label{fig:mg}
Influence of Ly$\beta$ fluorescence and choice of \MgII\  model atom on the
 \MgII\ K-band lines.}
\end{figure*}

{\bf{Iron.}} Two types of \FeII\ lines are found in the spectra
\footnote{The forbidden [\FeII] 1.677\um\ line present in the Pistol Star is
not included in our models}.
The first
are the strong semi-forbidden lines, including
z$^{4}$F$_{9/2}$-c$^{4}$F$_{9/2}$ 1.688~\um\ and
z$^{4}$F$_{3/2}$-c$^{4}$F$_{3/2}$~2.089~\um, that form in the outer wind and
have small oscillator strengths (gf$\sim$10$^{-5}$). The second are the
weak permitted (gf$\sim$1) lines connecting higher lying levels, such as the
4de$^{6}$G-5p$^{6}$F lines near 1.733~\um\ or 6p$^6$D-6s$^6$D at 2.109~\um,
that form much closer to the photosphere.

The permitted lines are more robust iron abundance indicators, having only
weak dependences on other parameters, such us turbulent velocity. The
strengths of the semi-forbidden lines depend on the accuracy of their
weak gf values, the mass loss rate and the
run of the iron ionization structure in the outer wind, which is
sensitive to the hydrogen ionization structure due to the strong coupling to
the Fe/H charge-exchange reactions. Since a change in the run of the 
clumping factor in the outer wind regions modifies the ratio of
recombinations/ionizations in hydrogen, the semi-forbidden lines are
diagnostic of the behavior of clumping there.
From Fig.~\ref{fig:pistol} it
can be seen that that our model is able to simultaneously reproduce both
sets of lines, providing constraints on both clumping and abundance.\footnote{We think
that the slight missmatch in the \FeII] 2.117~\um\ line in both objects is
related to the accuracy of the f value.}

\begin{figure}
\epsscale{0.90}     
\vspace{-.1cm}
\plotone{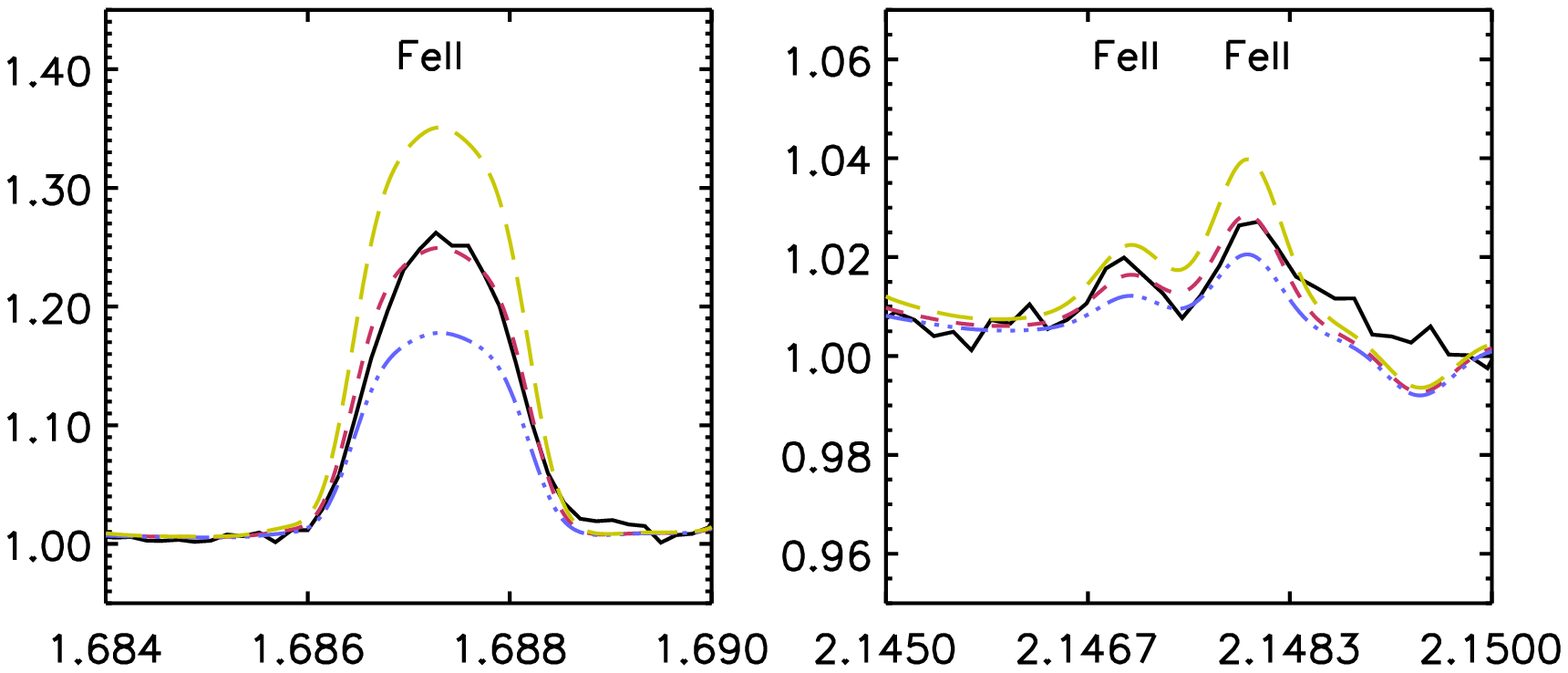}
\vspace{-.3cm}
\plotone{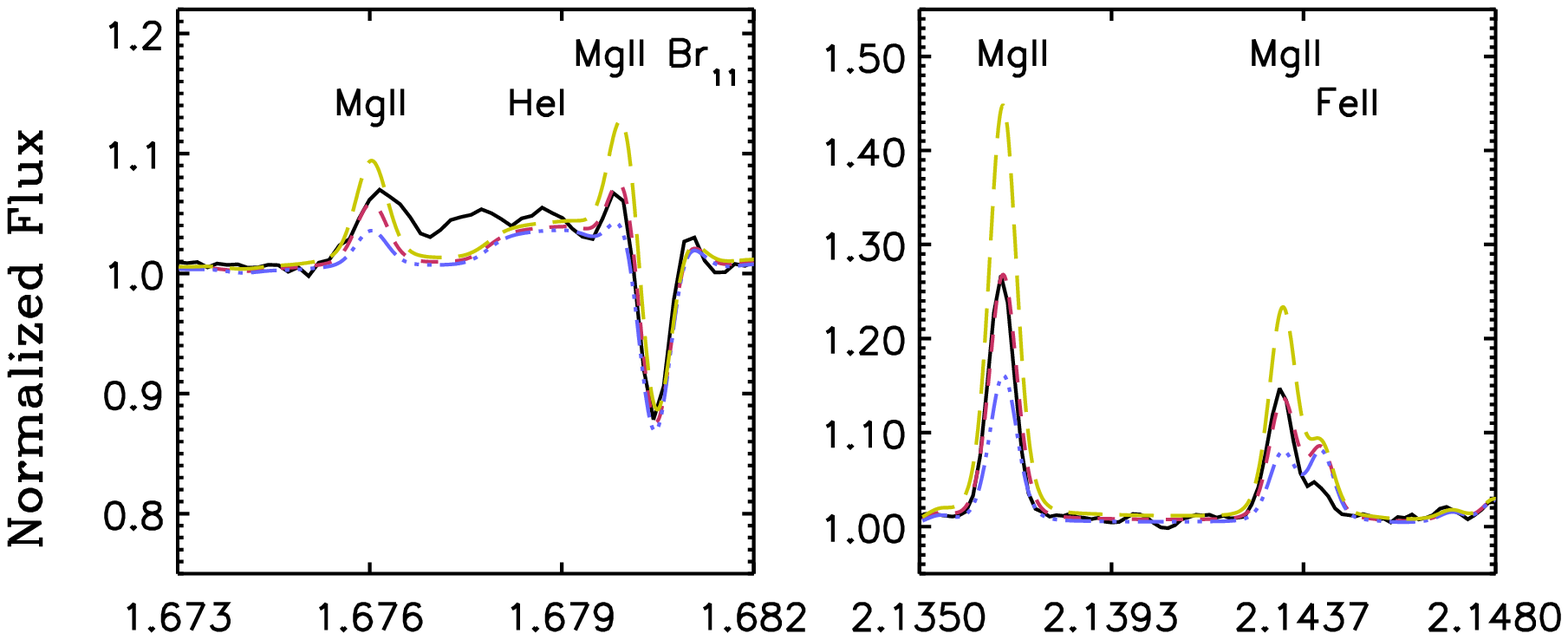}
\vspace{-.3cm}
\plotone{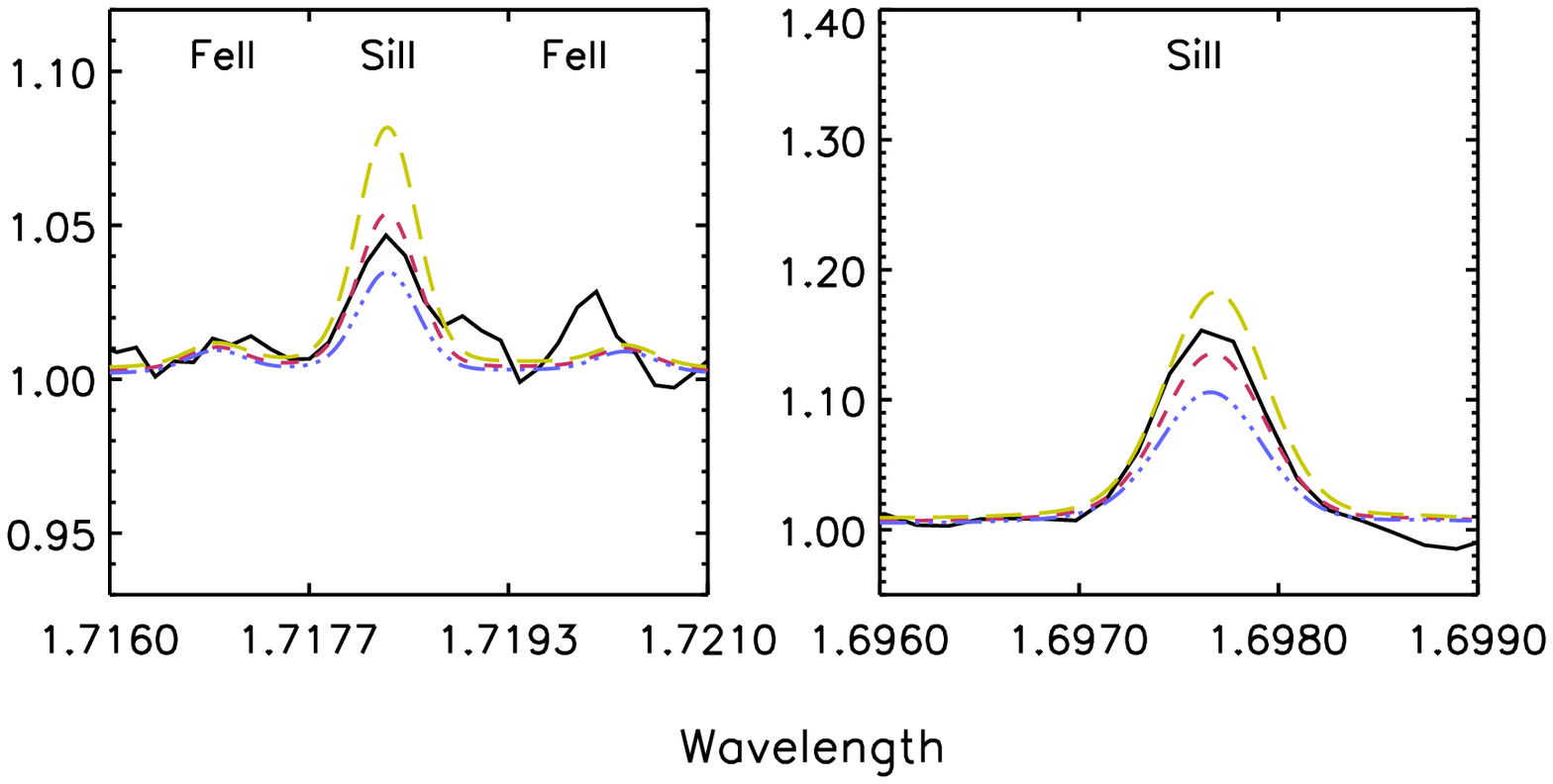}
\caption{\label{fig:lbv_metal}
Error estimates of Fe (upper-panel), Mg (middle-panel) and Si (lower-panel) abundandes. Dashed lines (red) correspond to
our best model fitting the observed (black-solid) diagnostic lines of 
\FMM362. Long-dashed (green) and dashed-dotted (blue) lines correspond to models
where individual metal abundances have been set to the derived upper and lower estimates
respectively.}
\end{figure}

We obtain roughly solar iron abundances for both LBVs, with $\pm$0.15dex as
plausible uncertainties (see Fig.~\ref{fig:lbv_metal}).
Our results are similar to A(Fe/H)=7.59 recently
derived by \citet{cun07} from their analysis of a sample of luminous cool
stars within 30~pc of the Galactic Center. Note in Table~1 that the Fe
abundance ratio has significant uncertainty due to the uncertainty in the Fe
abundance in the Sun.

{\bf{Magnesium.}} The strongest \MgII\ lines observed in the H and K bands
share the 5p$^2$P level. Those lines with it as the upper level, the
2.13/14~\um\ and 2.40/41~\um\ doublets (see Figs~\ref{fig:pistol} and
\ref{fig:362}) are much stronger than those with it as the lower level (H
band lines), revealing that pumping through the resonance 3s$^2$S-5p$^2$P
line must be a significant populator of the 5p$^2$P levels. Pumping through
the 3s$^2$S$[1/2]$-5p$^2$P$[3/2]$ 1025.968\AA\ transition is very efficient
due to Ly$\beta$ fluorescence. This was confirmed in models in which we
decoupled the 5p$^2$P$[3/2]$ and 5p$^2$P$[1/2]$ levels (see
Fig.~\ref{fig:mg}), resulting in \MgII\ 2.13/14~\um\ ratios much higher than
observed.

The relevance of this process can be easily followed in Fig.~\ref{fig:mg},
which displays the behavior of the doublet as a function of the choice of
the \MgII\ atom and the turbulent velocity. The latter refers to the fixed
Doppler width used in our models to compute the level populations. In the
left panel of Fig.~\ref{fig:mg} the levels are considered to be decoupled
(\ie\, the number of superlevels in the model atom, NS, is set to the total
number of levels in the full atom, NF). Because Ly$\beta$ lies closer to
3s$^2$S$[1/2]$-5p$^2$P$[3/2]$ ($\Delta$v=72\kms) than to
3s$^2$S$[1/2]$-5p$^2$P$[1/2]$ ($\Delta$v=114\kms), and because in these LBVS
the terminal velocities and wind densities determine the line formation
zones), only the 5p$^2$P$[3/2]$ is pumped thru fluorescence. Indeed, it can
be seen that as the turbulence velocity is increased, the overlap between
Ly$\beta$ and the \MgII\ line increases, as does the population of the
5p$^2$P$[3/2]$ level and the strength of the \MgII\ 2.13~\um\ line
increases, while the longer wavelength \MgII\ 2.14~\um\ line is unaffected.  
On the other hand, if the \MgII\ model atom has both levels combined into a
superlevel (NS$\neq$NF, Fig.~\ref{fig:mg}{-{right}}), the observed ratio is
reproduced. Furthermore, increasing the turbulent velocity, and hence the
effect of fluorescence, increases the pumping of both levels equally and
thus increases the strength of the doublet with a constant ratio between its
components. From Fig.~\ref{fig:mg} it can be seen that our assumed collision
coefficients connecting the \MgII\ 5p$^2$P$[1/2]$ and 5p$^2$P$[3/2]$ levels
may be too low. This comparison illustrates the importance of making the
correct choice of model atoms for quantitative spectroscopic analysis.

Due to fluorescence coupling, the \MgII\ K-band lines show a stronger
dependence on turbulent velocity than do the H-Band lines. We estimate about
twice solar Mg abundance and an associated uncertainty 
(see Fig.~\ref{fig:lbv_metal}) of about $\pm$0.25dex
(due to uncertainties related to the fluorescence contribution).

{\bf{Silicon.}} The \SiII\ doublet
5s$^{2}$S$_{1/2}$-5p$^{2}$P$_{3/2}$~1.691~\um\ and
5s$^{2}$S$_{1/2}$-5p$^{2}$P$_{1/2}$~1.698~\um\ constitutes a powerful
diagnostic tool, as it appears in emission for only a very narrow range of
stellar temperatures and wind density structures, indicating the presence of
amplified NLTE effects. However, since it forms at the base of the wind, its
strong dependence on the details of the velocity field there hinders a
precise silicon abundance determination.

Instead, we use the well-behaved recombination line \SiII\
3s$^2$6g$^2$G-3s$^2$5f$^2$F~at~1.718~\um\ which shows a stronger dependence
on the silicon abundance. Once again, a realistic mapping of full- to
super-levels in our model atom is required. From our model fits (see
Figs.~\ref{fig:pistol}~and~\ref{fig:362}) we derive roughly twice solar
abundance ($\pm$0.20~dex) for silicon in each LBV, similar to magnesium 
(see Fig.~\ref{fig:lbv_metal}).

{\bf{Other elements.}} One might expect a number of oxygen lines might be
detectable in infrared spectra of LBVs: i.e., strong \OI\ lines at
2.763~\um, 2.893~\um\ and 3.098~\um\ and weaker lines at 1.8243~\um,
3.661~\um\ and 3.946~\um. Several of these, but not all, are problematical
from ground-based observatories. Unfortunately our data set only encompasses
the \OI~1.745~\um\ line which is blended with a stronger \MgII\ line. Thus,
we defer an attempt to estimate the oxygen abundance until high resolution
observations of unblended lines can be obtained. Determining the oxygen
abundances in these objects will provide crucial constraints on their
evolutionary status.  Models show that when H/He$<$1.50 oxygen has reached
its maximum depletion within CNO equilibrium, while a significantly higher O
content should be present on the stellar surface for H/He values around 3.
The results in Table~1 then predict that the Pistol Star and \FMM362 have
different oxygen abundances. On the other hand, if an LBV has a H/He$<$1.50
but is still hydrogen rich, the oxygen abundance determination, expected to
be $\sim$0.04 of the original value, will provide a measure of the
metallicity of the natal cloud.  High resolution L-Band spectra of the
Pistol Star should be able to address this issue.

The only detected sodium lines are the well-known doublet at 2.206/9~\um,
from which we obtain a very high abundance, $\sim 20\times$solar. The strong
observed emission of this doublet in the K-band spectra of other LBVs has
been noted previously by \citet{hil98b}. Interestingly, our models display
only a minor dependence of these lines on clumping. On the other hand, the
strengths of the sodium lines might not indicate extraordinary sodium
abundance if the lines are produced by fluorescence of circumstellar
material, a component that we do not model.

{\section{Discussion.}}

Our results suggest solar Fe abundances and approximately twice-solar
$\alpha$-element abundances for the Quintuplet LBVs.  Presumably, these
abundances were the same in the gas that condensed to form these stars and
the other stars in the Quintuplet cluster and indeed in the whole of the
present-day Galactic center. The results can be discussed in the context of
similar measurements of Galactic center objects and with respect to the
trend one might expect if the region is an inward extension of the
disk or the bulge. In addition, the ratio of Fe to $\alpha$-elements might
be used to decipher the star formation history in the Galactic center.

\begin{deluxetable}{llllllll}     
\tabletypesize{\scriptsize}             
\tablecaption{Metallicity Estimates in the GC}
\tablehead{
\colhead{{}} &
\colhead{{A(Fe/H)}} &
\colhead{{A(O)}} &
\colhead{$\Sigma${CNO}}&
\colhead{{Z(N)}} &
\colhead{{A(Mg/H)}} &
\colhead{{A(Si/H)}} &
\colhead{{A(Ca/H)}}}
\startdata
Sun \\ 
\citet{and89}          & 7.67          & 8.93          & 1.6 & 1.5     & 7.58  & 7.55  &  6.36 \\
\citet{gre93}          & 7.50          & 8.87          & 1.4 & 1.3     & 7.58  & 7.55  &  6.36 \\
\citet{asp05}          & 7.45          & 8.66          & 0.82 &        & 7.53  & 7.51  &  6.31 \\
\\
Hot Stars\\
This work                   & 7.54$\pm0.15$ &               &              & & 7.85  & 7.84  &       \\
\citet{naj04}               &               &               &  & 1.57$\pm0.45$       &       &       &       \\
\citet{geb06}               &               & 8.91$\pm0.12$ &                &           &       &       &       \\
\citet{mar07a}              &               &               &  &  1.43$\pm0.42$             &       &       &       \\
\citet{mar07b}              &               &               &  & 1.80$\pm0.50$           &       &       &       \\
\\
Cool Stars\\
\citet{car00}               & 7.48$\pm0.13$ &               & 1.14$\pm0.55$ &&       &       &       \\
\citet{ram00}               & 7.61$\pm0.22$ &               &          &     &       &       &       \\
\citet{cun07}               & 7.59$\pm0.06$ & 9.04$\pm0.19$ &        &   & &       &  6.71$\pm0.14$  \\
\enddata

\tablecomments{Compilation of estimated stellar metal abundances at the GC.  
All abundances are given as A(X/H) except column~4 which displays the total
CNO mass percentage and column~5 displaying the nitrogen surface abundance
percentage Z(N). For the first two rows (solar values), column~5
gives the maximum nitrogen surface mass abundance, Z(N)$_{max}$,
reached during evolution according to the Geneva models.  For the other rows,
the average of the derived Z(N)  for the sample in each work is displayed.  
All studies of hot stars make use of CMFGEN, while those of cool stars
utilize \citet{ple92} models grid. The results of \citet{naj04} and
\citet{mar07b} are for stars in the Arches Cluster while those of
\citet{geb06} and \citet{mar07a} refer to  the Central Parsec
Cluster.  Star AFNW in \citet{mar07a} has not being considered due to 
the poorer S/N
and resolution quoted by the authors. The results of \citet{car00}
are for IRS7 while those of \citet{ram00} are for six M supergiants and
three giants. \citet{cun07} extends the sample of \citet{ram00} and 
also computes CNO and $\alpha$-elements abundances.
} 
\end{deluxetable}  


Table~2 displays a number of recent determinations of stellar 
metal abundances in the
Galactic Center together with the above mentioned three reference patterns
for solar abundances.  The values derived for Fe abundances in cool stars
agree with our result \citep{car00,ram97,ram99,ram00}.  \citet{cun07} find a
very narrow range of Fe abundances clustered around the solar value for a
population of cool stars in the central 30~pc. Of particular interest is
star VR5-7 from \citet{cun07} sample which is located in the Quintuplet
Cluster and shows A(Fe/H)=7.60 and A(Ca/H)=6.41

There are relatively few measurements of the $\alpha$-element abundances
([$\alpha$/Fe]) in GC stars.  \citet{naj04} find solar abundances (as
defined in this paper) for hot
stars in the Arches cluster based on the oxygen abundance and, to a lesser
degree, carbon abundance, and adopting the canonical solar value of
A(O/H)=8.93 \citep{and89}. Those estimates assume that nitrogen has reached
its maximum surface abundance value. Evolutionary models indicate that 95\%
of that value is already attained by the time that H/He$<$2 (by number).  
\citet{naj04} followed the metallicity patterns from the Geneva evolutionary
models and assumed no selective enrichment of CNO or $\alpha$-elements vs
Fe, in concluding that the stars in the Arches Cluster have solar
$\alpha$-element abundances. However, estimates of solar abundances have
varied considerably over the past 15 years
\citep[e.g.][, see also Table~2]{alle08}.
Thus, depending on
the assumed solar CNO composition, the derived nitrogen abundance by
\citet{naj04} could imply solar \citep[][]{and89}, 1.2~$\times$~solar
\citep[][]{gre93} or 2.0~$\times$~~solar \citep[][]{asp05} CNO composition.

Recently \citet{mar07a,mar07b} have analyzed a larger sample of hot stars in
the Arches and Central Parsec clusters and find similar results (see
Table~2). Interestingly, if one considers only the objects in \citet{mar07b}
with He/H$>0.1$ and those with Z(C)$<0.05$, \ie\ fulfilling the condition to
be close enough to Z(N)$_{max}$, the average value of Z(N) is 1.7.

\citet{geb06} estimate roughly solar oxygen abundance, A(O/H)=8.91, in IRS~8,
an OIf supergiant near the central parsec.  \citet{cun07} find
$<$A(O/H)$>$=9.04 ([0.37]) and $<$A(Fe/H)$>$=7.59 ([0.14]) for their sample
of cool stars, where the numbers in brackets are the ratio with respect to
the solar value in dex.  This implies [O/Fe]=0.22, i.e.\ a clear enhancement
over the solar ratio. Again, the \citet{cun07} measurements could be
interpreted as indicating solar ratios in O over Fe if the solar O abundance
in evolutionary models is used.  
It is crucial to have
accurate solar abundances, and that values used in stellar
evolution calculations should be consistent with these.
An excellent example is
attempting to determine whether the possible oxygen enhancement is due to a
top-heavy IMF favoring $\alpha$-elements vs Fe enrichment, or simply an
overall CNO and metal enhancement. Thus, taking the CNO abundances for the
GC objects from \citet{cun07} and assuming C/N equilibrium values one can
interprete their results either as solar CNO with mildly enhanced (30\%)
oxygen \citep{and89} or a clearly supersolar environment with a factor of
1.7 enhancement for C and N and 2.5 for oxygen \citep{asp05}.

Fortunately, there are other $\alpha$-elements whose adopted solar
abundances have suffered basically no major revision.  Thus, we believe that
the enhanced values obtained in this work for Mg and Si, roughly a factor of
two solar, together with the enhancement of Ca found by \citet{cun07}, are a
strong indication of the enrichment of $\alpha$-elements compared to Fe.

Our results run counter to the trend in the disk \citep{roll00,sma01,mar03},
and are more consistent with the values found for the bulge
\citep{fro99,fel00}. This may imply that the ISM in the disk does not extend
inward to the GC, so that material is dragged into the central molecular
zone from the bulge rather than from the disk. Another possibility is that
the GC stars are forming out of an ISM that has an enrichment history
distinctly different from that of the disk. At this point, further studies
of the $\alpha$-elements vs Fe would be useful. Future high S/N and high
resolution spectroscopy of the \OI\ lines in LBVs and K-band spectroscopy of
WNL stars in the same cluster (Najarro et al. in prep.) will provide two
independent
measurements of the original oxygen content, and thus set definite
constraints on metallicity.

The modest enrichment in $\alpha$-elements versus Fe that we find in the
two Quintuplet LBVs is consistent with a top-heavy IMF in the GC
\citep{fig99a}. In such a scenario, enhanced yields of $\alpha$-elements
compared to Fe are expected through a higher than average ratio of the
number of SNII vs SNIa events \citep{whe89,cun07}.

\acknowledgements

We thank Fabrice Martins and Katia Cuhna for usefull discussions.  F.~N. acknowledges
AYA2004-08271-C02-02 and AYA2007-67456-C02-02 grants. The material in this paper is based on work
supported by NASA under award NNG 05-GC37G, through the Long Term Space
Astrophysics program. TRG's research is supported by the Gemini Observatory,
which is operated by the Association of Universities for Research in
Astronomy, Inc., on behalf of the international Gemini partnership of
Argentina, Australia, Brazil, Canada, Chile, the United Kingdom, and the
United States of America. D. Figer is supported by a NYSTAR Faculty
Development Program grant.



\begin{thebibliography}{}
\bibitem[Afflerbach et al.(1997)]{affler97} Afflerbach, A.,  Churchwell, E.~B. \&\ Werner, M.W., 1997, \apj, 478, 190
\bibitem[Allende Prieto(2008)]{alle08} Allende Prieto, C.\ 2008, 14th Cambridge Workshop on Cool Stars, Stellar Systems, and the Sun, 384, 39 
\bibitem[Anders \& Grevesse(1989)]{and89} Anders, E. \&\ Grevesse, N., 1989, Geochimica et  Cosmochimica Acta, Vol. 53, 197
\bibitem[Asplund et al.(2005)]{asp05} Asplund, M., Grevesse, N., \& Sauval, A.~J.\ 2005, Cosmic Abundances as Records of Stellar Evolution and Nucleosynthesis, ASPC, 336, 25 
\bibitem[Carr et al.(2000)]{car00} Carr, J.S., Sellgren, K. \& Balachandran, S.C. 2000, \apj, 530, 307
\bibitem[Charbonnel et al.(1993)]{charb93} Charbonnel, C., Meynet, G., Maeder, A., Schaller, G., \& Schaerer, D.\ 1993, \aaps, 101, 415 
\bibitem[Cunha et al.(2007)]{cun07} Cunha, K., Sellgren, K., Smith, V.~V., Ramirez, S.~V., Blum, R.~D., \& Terndrup, D.~M.\ 2007, ArXiv e-prints, 707, arXiv:0707.2610 
\bibitem[Cotera et al.(1994)]{cot94} Cotera, A.~S., Erickson, E.~F., Allen, D.~A., Colgan, S.~W.~J., Simpson, J.~P., \& Burton, M.~G.\ 1994, NATO ASIC Proc.~445: The Nuclei of Normal Galaxies: Lessons from the Galactic Center, 217 
\bibitem[Crowther et al.(2006)]{cro06} Crowther, P.~A., Lennon, D.~J., \& Walborn, N.~R.\ 2006, \aap, 446, 279
\bibitem[Feltzing \& Gilmore(2000)]{fel00} Feltzing, S., \& Gilmore, G.\ 2000, \aap, 355, 949
\bibitem[Figer, McLean, \& Morris(1995)]{fig95} Figer, D.~F., McLean, I.~S., \& Morris, M.\ 1995, \apjl, 447, L29
\bibitem[Figer et al.(1998)]{fig98} Figer, D. F., Najarro, F., Morris, M., McLean, I. S., Geballe, T. R., Ghez, A. M., \& Langer, N. 1998, \apj, 506, 384
\bibitem[Figer, McLean, \& Najarro(1997)]{fig97} Figer, D.~F., McLean, I.~S., \& Najarro, F.\ 1997, \apj, 486, 420 
\bibitem[Figer et al.(1999a)]{fig99a} Figer, D.\ F., Kim, S.\ S., Morris, M., Serabyn, E., Rich, R.\ M., \& McLean, I.\ S.\ 1999a, \apj, 525, 750
\bibitem[Figer et al.(1999b)]{fig99b} Figer, D. F., McLean, I. S., \& Morris, M. 1999, \apj, 514, 202 
\bibitem[Figer et al.(1999c)]{fig99c} Figer, D.\ F., Morris, M., Geballe, T.\ R., Rich, R.\ M., Serabyn, E., McLean, I.\ S., Puetter, R.\ C., \& Yahil, A.\ 1999b, \apj, 525, 759 
\bibitem[Figer et al.(2002)]{fig02} Figer, D.~F., et al.\ 2002, \apj, 581, 258 
\bibitem[Frogel et al.(1999)]{fro99} Frogel, J.A., Tiede, G.P., \& Kuchinski, L.E.\ 1999, \aj, 117, 2296
\bibitem[Fuhrmann (1998)]{fuhr98} Fuhrmann, K., 1998, \aap, 338, 161
\bibitem[Geballe, Najarro, \& Figer (2000)]{geb00} Geballe, T.~R., Najarro, F., \& Figer, D.~F.\ 2000, \apjl, 530, L97 
\bibitem[Geballe et al.(2006)]{geb06} Geballe, T.~R., Najarro, F., Rigaut, F., \& Roy, J.-R.\ 2006, \apj, 652, 370 
\bibitem[Girardi et al.(2000)]{gir00} Girardi, L., Bressan, A., Bertelli, G., \& Chiosi, C.\ 2000, \aaps, 141, 371 
\bibitem[Glass et al.(1987)]{gla87} Glass, I.~S., Catchpole, R.~M., \& Whitelock, P.~A.\ 1987, \mnras, 227, 373 
\bibitem[Glass et al.(1990)]{gla90} Glass, I.~S., Moneti, A., \& Moorwood, A.~F.~M.\ 1990, \mnras, 242, 55P 
\bibitem[Glass et al.(1999)]{gla99} Glass, I.~S., Matsumoto, S., Carter, B.~S., \& Sekiguchi, K.\ 1999, \mnras, 304, L10
\bibitem[Grevesse \& Noels(1993)]{gre93} Grevesse, N., \& Noels, A.\ 1993, Origin and evolution of the elements: proceedings of a symposium in honour of H.~Reeves, held in Paris, June 22-25, 1992.~Edited by N.~Prantzos, E.~Vangioni-Flam and M.~Casse.~Published by Cambridge University Press, Cambridge, England, 1993, p.14, 14 
\bibitem[Grevesse \& Sauval(1998)]{gre98} Grevesse, N., \& Sauval, A.~J.\ 1998, Space Science Reviews, 85, 161 
\bibitem[Groh et al.(2006)]{gro06} Groh, J.~H., Hillier, D.~J., \& Damineli, A.\ 2006, \apjl, 638, L33
\bibitem[Herald et al.(2001)]{her01} Herald, J.~E., Hillier, D.~J., \& Schulte-Ladbeck, R.~E.\ 2001, \apj, 548, 932 
\bibitem[Hillier \& Miller(1998)]{hil98} Hillier, D.~J.~\& Miller, D.~L.\ 1998, \apj, 496, 407
\bibitem[Hillier et al.(1998)]{hil98b}  Hillier, D.J., Crowther, P.A., Najarro, F., Fullerton, A.W., 1998b, \aap, 340, 483
\bibitem[Hillier \& Miller(1999)]{hil99} Hillier, D.~J.~\& Miller, D.~L.\ 1999, \apj, 519, 354
\bibitem[Hillier et al.(2001)]{hil01} Hillier, D.~J., Davidson, K., Ishibashi, K., \& Gull, T.\ 2001, \apj, 553, 837
\bibitem[Iglesias \& Rogers(1996)]{igl96} Iglesias, C.~A., \& Rogers, F.~J.\ 1996, \apj, 464, 943 
\bibitem[Kennicutt et. al.(2003)]{kenni03} Kennicutt, R.C., Bresolin, F., Garnett, D.R. 2003, \apj, 591, 544
\bibitem[Kudritzki \& Puls(2000)]{kud00} Kudritzki, R.-P., \& Puls, J.\ 2000, \araa, 38, 613 
\bibitem[Luck et al.(2006)]{luck06} Luck, R.~E., Kovtyukh, V.~V., \& Andrievsky, S.~M.\ 2006, \aj, 132, 902 
\bibitem[Maciel \&\ Quireza (1999)]{mac99} Maciel, W.J., \&\ Quireza, C. \ 1999, \aap, 345, 629
\bibitem[Maeda et al.(2000)]{mae00} Maeda, Y., et. al.,  2002, \apj, 570, 671
\bibitem[Mart{\'{i}}n-Hern\'andez et al.(2003)]{mar03} Mart{\'{i}}n-Hern\'andez, N.L., van der Hulst, J.M., \&\ Tielens, A.G.G.M.,  2003, \aap, 407, 957
\bibitem[Martins et al.(2007a)]{mar07a} Martins, F., Genzel, R., Hillier, D.~J., Eisenhauer, F., Paumard, T., Gillessen, S., Ott, T., \& Trippe, S.\ 2007, \aap, 468, 233
\bibitem[Martins et al.(2007b)]{mar07b} Martins, F., Hillier, D.~J., et al., 2007, \aap, (submitted)
\bibitem[Meynet et al.(1994)]{mey94} Meynet, G., Maeder, A., Schaller, G., Schaerer, D., \& Charbonnel, C. 1994, \aap\ Supp., 103, 97
\bibitem[Meynet \& Maeder(2003)]{mey03} Meynet, G., \& Maeder, A.\ 2003, \aap, 404, 975 
\bibitem[Meynet \& Maeder(2005)]{mey05} Meynet, G., \& Maeder, A.\ 2005, \aap, 429, 581
\bibitem[Moneti et al.(1994)]{mon94} Moneti, A., Glass, I. S. \& Moorwood, A. F. M. 1994, \mnras, 268, 194
\bibitem[Nagata et al.(1990)]{nag90} Nagata, T., Woodward, C.~E., Shure, M., Pipher, J.~L., \& Okuda, H.\ 1990, \apj, 351, 83 
\bibitem[Najarro et al.(1999)]{naj99} Najarro, F., Hillier, D.~J., Figer, D.~F., \& Geballe, T.~R.\ 1999, The Central Parsecs of the Galaxy, ASPC, 186, 340 
\bibitem[Najarro (2001)]{naj01} Najarro, F.\ 2001, P Cygni  2000: 400 Years of Progress, ASPC, 233, 133 
\bibitem[Najarro et al.(2004)]{naj04} Najarro, F., Figer, D. F., Hillier, D. J., \& Kudritzki, R. P.\ 2004 (Paper I), \apjl, 611, L105 
\bibitem[Nugis et al.(1998)]{nug98} Nugis, T., Crowther, P.~A., \& Willis, A.~J.\ 1998, \aap, 333, 956
\bibitem[Okuda et al.(1990)]{oku90} Okuda, H., et al.\ 1990, \apj, 351, 89 
\bibitem[Pinsonneault \& Delahaye(2006)]{pin06} Pinsonneault, M.~H., \& Delahaye, F.\ 2006, ArXiv Astrophysics e-prints, arXiv:astro-ph/0606077 
\bibitem[Plez(1992)]{ple92} Plez, B.\ 1992, \aaps, 94, 527
\bibitem[Puls et al.(2006)]{pul06} Puls, J., Markova, N., Scuderi, S., Stanghellini, C., Taranova, O.~G., Burnley, A.~W., \& Howarth, I.~D.\ 2006, \aap, 454, 625
\bibitem[Ramirez et al.(1997)]{ram97} Ramirez, S.V., Carr, J.S., Balachandran, S., Blum, R., \& Terndrup, D.M.\ 1997, in IAU Symp. 184, The Central Regions of the Galaxy and Galaxies, ed. Y. Sofue (Dordrecht: Kluwer), 28
\bibitem[Ramirez et al.(1999)]{ram99} Ramirez, S.V., Sellgren, K., Carr, J.S., Balachandran, S., Blum, R., \& Terndrup, D.M.\ 1999, ASP Conf.~Ser.~186: The Central Parsecs of the Galaxy
\bibitem[Ramirez et al.(2000)]{ram00} Ramirez, S.V., Sellgren, K., Carr, J.S., Balachandran, S., Blum, R., Terndrup, D.M., \&\ Steed, A., 2000, \apj, 537,205
\bibitem[Rolleston et. al.(2000)]{roll00} Rolleston, W.R., Smartt, S.J., Dufton, P.L., \&\ Ryans, R.S.I., 2000, \aap, 363, 537
\bibitem[Runacres \& Owocki(2002)]{run02} Runacres, M.~C.,\& Owocki, S.~P.\ 2002, \aap, 381, 1015 
\bibitem[Rudolph et al.(2006)]{rud06} Rudolph, A.~L., Fich, M., Bell, G.~R., Norsen, T., Simpson, J.~P., Haas, M.~R., \& Erickson, E.~F.\ 2006, \apjs, 162, 346 
\bibitem[Salasnich et al.(2000)]{sal00} Salasnich, B., Girardi, L., Weiss, A., \& Chiosi, C.\ 2000, \aap, 361, 1023 
\bibitem[Schaller et al.(1992)]{scha92} Schaller, G., Schaerer, D., Meynet, G., \& Maeder, A.\ 1992, \aaps, 96, 269
\bibitem[Shields \&\ Ferland (1994)]{shi94}Shields, J.C.,  \&\ Ferland, G.J., 1994, \apj, 430, 236
\bibitem[Smartt et. al.(2001)]{sma01} Smartt, S.J., Venn, K.A.,  Dufton, P.L., Lennon, D.J., Rolleston, W.R.,  \&\ Keenan, F.P., 2001, \aap, 367, 86
\bibitem[Stahl et al.(2001)]{sta01} Stahl, O., Jankovics, I., Kov{\'a}cs, J., Wolf, B., Schmutz, W., Kaufer, A., Rivinius, T., \& Szeifert, T.\ 2001, \aap, 375, 54 
\bibitem[Urbaneja et al.(2005)]{urba05} Urbaneja, M.A., Herrero, A., Bresolin, F., Kudritzki, R.P., Gieren, W., Puls, J., Przybilla, N., Najarro, F., Pietrzynski, G., 2005, \apj, 622, 862
\bibitem[Vink \& de Koter(2002)]{vin02} Vink, J.~S., \& de Koter, A.\ 2002, \aap, 393, 543 
\bibitem[Wheeler et al.(1989)]{whe89} Wheeler, J.~C., Sneden, C., \& Truran, J.~W., Jr.\ 1989, \araa, 27, 279
\end{thebibliography}
\end{document}